# A knockoff filter for high-dimensional selective inference

Rina Foygel Barber* and Emmanuel J. Candès†

## Abstract

This paper develops a framework for testing for associations in a possibly high-dimensional linear model where the number of features/variables may far exceed the number of observational units. In this framework, the observations are split into two groups, where the first group is used to screen for a set of potentially relevant variables, whereas the second is used for inference over this reduced set of variables; we also develop strategies for leveraging information from the first part of the data at the inference step for greater power. In our work, the inferential step is carried out by applying the recently introduced knockoff filter, which creates a knockoff copy—a fake variable serving as a control—for each screened variable. We prove that this procedure controls the *directional* false discovery rate (FDR) in the reduced model controlling for all screened variables; this says that our high-dimensional knockoff procedure 'discovers' important variables as well as the directions (signs) of their effects, in such a way that the expected proportion of wrongly chosen signs is below the user-specified level (thereby controlling a notion of Type S error averaged over the selected set). This result is non-asymptotic, and holds for any distribution of the original features and any values of the unknown regression coefficients, so that inference is not calibrated under hypothesized values of the effect sizes. We demonstrate the performance of our general and flexible approach through numerical studies, showing more power than existing alternatives. Finally, we apply our method to a genome-wide association study to find locations on the genome that are possibly associated with a continuous phenotype.

## 1 Introduction

Many modern studies in the sciences take the following form: collect a large amount of data, and then ask which of the potentially many measured variables are possibly associated with a response of interest. A typical example would be a genome-wide association study (GWAS), where one wishes to 'discover' which genetic variants are possibly associated with a trait. In such studies, it is common to read genetic variants using single-nucleotide polymorphism (SNP) arrays and then look for associations between the trait and the hundreds of thousands of SNPs. In these types of studies and others, we typically have many more variables or features (SNPs) than observations (individuals in the study).

Finding statistically significant associations between a response and a large set of potentially explanatory variables requires the specification of a statistical model; in this work, we consider the classical linear regression model, which takes the form

$$Y = \sum_{j=1}^{p} \beta_j X_j + \epsilon, \qquad (1)$$

where $Y$ is the response variable of interest, the $X_j$'s are the explanatory variables, and $\epsilon$ is a (stochastic) error term. In the GWAS example, $Y$ may be the level of HDL cholesterol and $X_j$ the number of recessive alleles at a given location on the genome. In such applications it is important to keep in mind that the linear relationship (1) is just an approximation of a more complicated and unknown model.

---

*Department of Statistics, University of Chicago
†Departments of Statistics and Mathematics, Stanford University

In classical statistics, when we speak of finding associations, we think of testing each of the $p$ hypotheses $H_j : \beta_j = 0$, and one can see that there are a few complications with this viewpoint. The first is that for fixed designs, the linear model (1) with $p$ parameters is not identifiable whenever the number $n$ of samples is smaller than this number $p$ of features. Formally, testing an unidentifiable model may seem like a problematic proposition. A second complication is that in our GWAS example, it may very well be the case that causal mutations affecting levels of HDL cholesterol, for instance, have not been typed (i.e. the array of SNPs is a subsample of the genome, and does not include the causal mutation). In such studies one is, therefore, led to search for indirect effects. In other words, while the causal factors have not been measured, we nevertheless hope to have measured variables that are sufficiently close or correlated to be picked up by association. In these settings, it is understood that even if there is a true sparse 'causal' model, we may end up working with an approximate and non-sparse statistical model of the form (1) in which few, if any, of the features have precisely zero effect (i.e. for most features $j$, $\beta_j = 0$ is not exactly true); this phenomenon occurs because the causal effects have been folded onto the observed variables. There are two consequences of this: (1) sparsity assumptions, which are traditionally assumed in the literature to relieve ourselves from the identifiability problem, are not applicable[1] and (2) when few or none of the regression coefficients take on the value zero, controlling a classical Type I error typically makes little sense since nearly any selected feature is technically a true positive, and we need a different measure of performance to capture the idea that we would like to correctly identify the meaningful large effects while screening out features with near-zero effects. To summarize our discussion, we have at least two problems:

- Lack of identifiability (at least when $p > n$).

- Possibly correlated features and lack of model sparsity so that few, if any, of the null hypotheses actually hold.

These complications should not occlude the relatively simple goal of a statistical selection procedure, which in the GWAS context can be stated as follows: *understanding that SNPs may be proxies for causal mutations, we would like to find as many proxies or locations on the genome without too many false positives—a false positive being a reported location (SNP) with no causal mutation sitting nearby*. Being guaranteed a sufficiently small fraction of false positives would assure the scientist that most of her discoveries are indeed true and, in some sense, replicable. We are thus interested in reliable methods for finding the large effects, those SNPs with large coefficient magnitudes. This paper develops a framework and methods for getting close to this goal. At a high level, we propose a two-stage procedure: first, using a portion of the data set to screen for a large set of potentially relevant features (which will likely include a high proportion of false positives), and second, applying the recently developed *knockoff filter* to the second portion of the data in order to perform inference within the submodel selected at the screening stage.

## 1.1  Errors of Type S and directional FDR

In a series of papers, Gelman and his collaborators introduce a useful measure they call *error of Type S* (S stands for sign). In their work [11], they argue that "in classical statistics, the significance of comparisons (e. g. , $\theta_1 - \theta_2$) is calibrated using Type I error rate, relying on the assumption that the true difference is zero, which makes no sense in many applications." They then propose "a more relevant framework in which a true comparison can be positive or negative, and, based on the data, you can state "$\theta_1 > \theta_2$ with confidence", "$\theta_2 > \theta_1$ with confidence", or "no claim with confidence"." In this framework, a Type S error (sometimes called a Type III error in other works) occurs when we claim with confidence that the comparison goes one way when, in fact, it goes the other way. This echoes an earlier point made by Tukey and we quote from [33]: "All we know about the world teaches us that the effects of $A$ and $B$ are always different—in some decimal place—for any $A$ and $B$. Thus asking 'Are the effects different' is foolish. What we should be answering first is 'Can we tell the direction in which the effects of $A$ differ from the effects of $B$?'." We refer to [33] as well as [19] and [5] for additional discussion.

---

[1]The literature also typically assumes low correlations between variables, which is also not applicable in the GWAS example.

33   02/2016; Revised 09/2017 and 05/2018
This point of view is extremely relevant to our (possibly high-dimensional) regression problem and our goal of finding the important effects. When we choose to report a variable, we should be able to state with confidence its direction of effect, i.e. whether $\beta_j > 0$ or $\beta_j < 0$. To press this point, we would surely have little faith in a procedure that would control errors of Type I but would not be able to tell the directions of the effects. Another useful aspect of Type S errors is that small effects $\beta_j \approx 0$ are generally those for which we cannot have much certainty about their direction, since the size of the effect is comparable to or below the noise level of our estimate. Therefore, if we were concerned by signed errors, we would probably not report these effects as discoveries (even if we believe that no effect is exactly zero)—arguably a wise thing to do.

To measure our success at selecting only those effects that are large enough to be meaningfully distinguished from noise, we might like to control the mixed directional false discovery rate (FDR$_{\text{dir}}$) [27, 6] defined as follows: letting $\widehat{S} \subset \{1, \ldots, p\}$ be the set of selected features together with estimates $\widehat{\text{sign}}_j \in \{\pm 1\}$ of the direction of effect,

$$\text{FDR}_{\text{dir}} = \mathbb{E}\left[\text{FDP}_{\text{dir}}\right],$$

where the mixed directional false discovery proportion (FDP$_{\text{dir}}$) is given by

$$\text{FDP}_{\text{dir}} = \frac{\left|\left\{j \in \widehat{S} : \widehat{\text{sign}}_j \neq \text{sign}(\beta_j)\right\}\right|}{|\widehat{S}| \vee 1} \quad (2)$$

with the convention that $\text{sign}(0) = 0$. This definition subsumes Type I and Type S errors: a false discovery or error occurs either when a zero effect is selected, or when a nonzero effect is selected but with the incorrect sign. (In the denominator, $|\widehat{S}| \vee 1$ denotes $\max\{|\widehat{S}|, 1\}$; this maximum is taken so that if we make zero discoveries, then the FDP is calculated as zero.) The term "mixed" comes from this combination of two error types; for our purposes, we consider declaring that $\beta_j > 0$ to be a sign error whether the truth is that $\beta_j = 0$ or $\beta_j < 0$, and will not distinguish between these two types of errors, and so we will drop the term "mixed" and refer to it simply as the "directional FDR". The directional FDP is then the total number of errors of this (combined) type averaged over the selected set. In regression settings in which many of the $\beta_j$'s are approximately zero but perhaps not exactly zero, controlling the directional FDR may be more appropriate than the "classical" FDR (which counts Type I errors only):

$$\text{FDR} = \mathbb{E}\left[\text{FDP}\right], \quad \text{FDP} = \frac{\left|\left\{j \in \widehat{S} \text{ and } \beta_j = 0\right\}\right|}{|\widehat{S}| \vee 1}.$$

Note that controlling the directional FDR has a different flavor than classical FDR, since paraphrasing Gelman et al., the significance is not calibrated using Type I errors which implicitly assumes that some of the $\beta_j$'s are zero. In contrast, we are after a form of inference holding no matter the values of the $\beta_j$'s. In classical settings with independent statistics, where $y_i \stackrel{\text{ind}}{\sim} \mathcal{N}(\mu_i, 1)$, say, and we wish to test for the means, [6] establishes that the Benjamini-Hochberg procedure [4] achieves directional FDR control.

In earlier work [1] we introduced *the knockoff filter*, a new variable selection procedure, which rigorously controls the classical FDR, in the low-dimensional setting where $p \leq n$. It should be clear that the equivalent notion (2) associated with Type S errors is in principle more difficult to control since by definition we have

$$\text{FDR}_{\text{dir}} \geq \text{FDR},$$

due to the fact that the classical FDR does not record an error when the $j$th feature is correctly included in the model but with the incorrect sign, while the directional FDR does. A first contribution of this paper is to show that the knockoff filter offers more than advertised in our earlier work [1]: indeed, we will find that this filter, with no modifications, actually controls the directional FDR as well as the FDR. In particular, this implies that the knockoff filter is applicable in settings where we do not expect any sparsity but nonetheless wish to test whether our conclusions regarding effect signs are reliable. This addresses one of the complications discussed earlier.

## 1.2 High dimensionality

To address the other complication, namely, the high-dimensionality issue ($p > n$), we propose the following general strategy:

- *Screening step.* In a first step, screen all the features $\{X_j\}$, $j = 1, \ldots, p$, as to identify a set $\widehat{S}_0 \subset \{1, \ldots, p\}$ of potentially relevant features with $|\widehat{S}_0| < n$. In this paper, we shall regard this step as being quite liberal in the sense that it typically produces a long list, hopefully containing most of the important features (with large effect sizes) but also possibly many features with zero effects $\beta_j = 0$ or nearly vanishing effects $\beta_j \approx 0$.

- *Inference/selection step.* The screening step yields a reduced model

$$Y = \sum_{j \in \widehat{S}_0} \beta_j^{\text{partial}} X_j + \epsilon, \tag{3}$$

where we use the notation $\beta^{\text{partial}}$ to indicate that both the definition and the meaning of the regression coefficients has changed (see below). Then test for associations in this reduced model by controlling the directional FDR.

In our GWAS example, this means that we would first screen for promising SNPs, and then extract from this set a final and smaller list of SNPs we are confident are associated with the phenotype under study (recall that reporting a SNP only means that we are confident a mutation lies nearby). This approach—a screening step, followed by inference on the screened submodel—has been studied by many researchers; see [37, 35] for a non-exhaustive list of important works among those lines.

We would like to emphasize the importance of considering the directional FDR rather than the classical (unsigned) FDR in the framework of a screening step followed by an inference step. Imagine that our sampled observations $\mathbf{y} \in \mathbb{R}^n$ have mean $\mathbf{X}\boldsymbol{\beta}$ and uncorrelated errors. In the reduced model, we wish to provide inference about the *partial regression coefficients*, $\boldsymbol{\beta}^{\text{partial}} \in \mathbb{R}^{|\widehat{S}_0|}$, when we regress the response $\mathbf{y}$ onto the features $\mathbf{X}_{\widehat{S}_0}$ only. Even if the original coefficient vector $\boldsymbol{\beta}$ for the full regression model is exactly sparse with many entries exactly equal to zero, the vector of partial regression coefficients $\boldsymbol{\beta}^{\text{partial}}$ is nonetheless likely to be dense unless there is special structure within $\mathbf{X}$ (e.g. if columns of $\mathbf{X}$ are orthogonal to each other).

Now, consider a feature $\mathbf{X}_j$ not appearing in the original full model, i.e. $\beta_j = 0$. In the partial regression, we may find one of two scenarios:

- The coefficient in the partial regression may be large, that is, $\beta_j^{\text{partial}} \not\approx 0$. This typically occurs if $\mathbf{X}_j$ is highly correlated with some strong signal $\mathbf{X}_k$ that was missed by the screening step. (For instance, we may have missed a causal SNP but included a close neighbor, instead, in the screened set.) Here, we would generally prefer to include $\mathbf{X}_j$ in the selected model in any case, since it is a good proxy for the missed relevant feature $\mathbf{X}_k$.

- Alternately, the coefficient might remain near zero, $\beta_j^{\text{partial}} \approx 0$. This is likely whenever $\mathbf{X}_j$ is not a proxy for any missed feature. Here, the sign of $\beta_j^{\text{partial}}$ cannot be estimated with much certainty, and so with directional FDR control, we are likely to exclude $\mathbf{X}_j$ from our final model.

These considerations hopefully make clear that careful screening followed by directional FDR control might yield a valuable selection procedure, which in our GWAS example would likely be able to report SNPs (locations on the genome) associated with a trait while controlling a very natural notion of false positive rate. On the other hand, a setting where this framework does not apply is that of causal inference, in which one would like to report true causal effects and would prefer not to select variables that, due to correlation, act as proxies to the true causal features.

## 1.3 Comparison with other works

There is of course a vast literature on inference in the regression setting. When $n > p$, we have available the sampling distribution of least-squares estimates, which makes the construction of p-values and confidence intervals possible, although selective inference procedures controlling the FDR and similar criteria are more tricky because of the correlations between such test statistics. For an overview of some of the work when $n > p$, we refer the reader to the companion article [1].

There has also been much recent work on the high-dimensional setting $p > n$, which is the focus of our work here. One type of approach to high-dimensional inference is to avoid the identifiability problem by making assumptions on the data distribution such as (1) a highly sparse coefficient vector $\boldsymbol{\beta}$, and/or (2) a "beta-min" condition which requires nonzero entries of $\boldsymbol{\beta}$ to not be too close to zero, and/or (3) assumptions on the design matrix such as low-pairwise correlations or placing a random distribution on the features. These types of conditions ensure that, with high probability (asymptotically), a sure screening property will hold: that is, at any place in the method requiring a model selection step, with high probability all the relevant features (i.e. all $\mathbf{X}_j$ with $\beta_j \neq 0$) will be selected. Under these conditions, it is possible to produce an asymptotically normal estimator for each coefficient $\beta_j$ which can then yield a p-value or confidence interval which is asymptotically valid. Methods in this category include the work of Voorman et al. [34], Belloni et al. [3], and Huang et al. [14]. A different type of result, relying on rapid decay of the error in estimating $\boldsymbol{\beta}$ in an asymptotic setting instead of requiring a sure screening property, is the debiasing work of Zhang and Zhang [40] and Javanmard and Montanari [18]; see also Voorman et al. [34] for a different but related approach. Lockhart et al. [22] propose a sequential testing procedure, which moves along the Lasso path as variables are added into the model, under the assumption that the signals appear before the false positives along the Lasso path (which holds with high probability asymptotically under similar assumptions, i.e.. highly sparse $\boldsymbol{\beta}$, the beta-min condition, and a well-behaved design matrix). To contrast these lines of work with our method presented here, empirically we observe reliable performance in terms of FDR control at even relatively small sample sizes, and our theoretical framework avoids strong assumptions on signal strength or sparsity level or the design matrix.

In contrast with the asymptotic setting described above, other approaches focus on finite-sample p-values and confidence intervals which do not rely on "high probability" selection events but instead are calculated by conditioning on the screened or selected model. Lee et al. [20] develops exact post-selection inference for the Lasso when applied with a fixed penalty parameter, with methodology that extends to other model selection procedures which can be expressed as linear constraints on $\mathbf{y}$; Tian et al. [30] extend this methodology to the setting of the square-root-Lasso, where the noise level $\sigma$ is no longer assumed to be known. In contrast, Tibshirani et al. [32] propose methods for sequential selection procedures, such as the path of selected models obtained by tuning the Lasso penalty parameter; in this setting, FDR control can be obtained with sequential hypothesis testing procedures e.g. [13], and the tests can be viewed as asking whether the most recent added feature has a true effect in the partial model fitted so far. Fithian et al. [10] develops general theory for this type of inference after conditioning on (sequential or nonsequential) selection events, termed "data carving". In contrast to these methods, which perform inference on the outcome of specific model selection methods (such as the Lasso), the knockoff framework gives the flexibility of choosing (nearly) any test statistics for model selection.

In earlier work, Leeb and Pötscher [21] studied the impossibility of characterizing the (unconditional) distribution of post-selection estimators, and the PoSI ("post-selection inference") framework of Berk et al. [7] proposed handling the range of possible selected models by taking a maximum over the set of possibilities. In practice, this type of method may have high power when considering extremely small submodels, but lose power when the selected model being considered grows larger.

Finally, we have called attention on Type S errors as a useful error metric for controlled variable selection. (Methods such as Lee et al. [20]'s post-selection inference for the Lasso can provide p-values testing a one-sided (i.e. signed) hypothesis, and can therefore be used to control directional FDR, but since this literature focuses on inference for individual hypotheses rather than FDR control, this distinction is not discussed explicitly.) We believe this is a correct framework for the model selection problem in the high-dimensional setting, where strict sparsity assumptions may be violated whenever a reduced model is considered.

## 2 Knockoffs

Since this paper builds upon and extends the knockoff methodology introduced in [1], we first review the essentials of this method. The literature offers many tools for estimating a sparse linear model in a high-dimensional setting, i.e. $\mathbf{y} \sim \mathcal{N}(\mathbf{X}\boldsymbol{\beta}, \sigma^2 \mathbf{I})$ where $\mathbf{X} \in \mathbb{R}^{n \times p}$ is a fixed design matrix and $\boldsymbol{\beta}$ is believed to be sparse.[2] For instance, we may use a forward stepwise selection method such as orthogonal matching pursuit (OMP) [24], where features are sequentially added to the model, choosing at each step the feature $\mathbf{X}_j$ that offers the greatest reduction in the size of the residual. Alternately, we might choose a convex optimization method such as the Lasso [31], where

$$\hat{\boldsymbol{\beta}}_{\text{Lasso}}(\lambda) = \arg\min_{\mathbf{b}\in\mathbb{R}^p}\left\{\tfrac{1}{2}\|\mathbf{y}-\mathbf{Xb}\|_2^2 + \lambda\|\mathbf{b}\|_1\right\} \tag{4}$$

---

[2] Recent work by the second author and collaborators [8] treats this problem in the alternate setting where the design matrix $\mathbf{X}$ is random, with a known distribution.

A critical question is this: can we estimate the number of false discoveries in the resulting coefficient vector $\hat{\boldsymbol{\beta}}_{\text{OMP}}$ or $\hat{\boldsymbol{\beta}}_{\text{Lasso}}$? The knockoff filter provides answers to questions of this kind by constructing a knockoff companion for each variable, which acts as a *control* for the possibility that we choose variable $j$ when it is in fact null, i.e. $\beta_j = 0$.

## 2.1 Knockoff variables

For each feature $\mathbf{X}_j$, a knockoff copy $\widetilde{\mathbf{X}}_j$ is constructed to satisfy

$$\begin{bmatrix}\mathbf{X} & \widetilde{\mathbf{X}}\end{bmatrix}^\top\begin{bmatrix}\mathbf{X} & \widetilde{\mathbf{X}}\end{bmatrix} = \begin{bmatrix}\mathbf{X}^\top\mathbf{X} & \mathbf{X}^\top\widetilde{\mathbf{X}} \\ \widetilde{\mathbf{X}}^\top\mathbf{X} & \widetilde{\mathbf{X}}^\top\widetilde{\mathbf{X}}\end{bmatrix} = \begin{bmatrix}\boldsymbol{\Sigma} & \boldsymbol{\Sigma}-\text{diag}\{\mathbf{s}\} \\ \boldsymbol{\Sigma}-\text{diag}\{\mathbf{s}\} & \boldsymbol{\Sigma}\end{bmatrix} \triangleq \mathbf{G}, \tag{5}$$

for some vector $\mathbf{s}\geq \mathbf{0}$.[3] Thus the knockoffs exhibit the same correlation structure as the original variables, and the cross-correlations are also preserved, in the sense that $\mathbf{X}_j^\top\mathbf{X}_k = \widetilde{\mathbf{X}}_j^\top\mathbf{X}_k$ for all $j\neq k$. While past methods in the statistics literature have considered adding fake variables in order to control false positives in regression constructed through other methods, such as by generating independent features or by permuting entries of existing features (see e.g. Miller [23], Wu et al. [38]), the knockoffs construction is unique in that it yields an exchangeability property: for any subset $S$ of nulls, we have

$$\begin{bmatrix}\mathbf{X} & \widetilde{\mathbf{X}}\end{bmatrix}^\top_{\text{swap}(S)}\mathbf{y} \stackrel{d}{=} \begin{bmatrix}\mathbf{X} & \widetilde{\mathbf{X}}\end{bmatrix}^\top\mathbf{y}, \tag{6}$$

where the matrix $\begin{bmatrix}\mathbf{X} & \widetilde{\mathbf{X}}\end{bmatrix}_{\text{swap}(S)}$ is obtained from $\begin{bmatrix}\mathbf{X} & \widetilde{\mathbf{X}}\end{bmatrix}$ by swapping the columns $\mathbf{X}_j$ and $\widetilde{\mathbf{X}}_j$ for each $j\in S$. For example, (6) implies that $|\mathbf{X}_j^\top\mathbf{y}|$ is equally likely to be larger or smaller than $|\widetilde{\mathbf{X}}_j^\top\mathbf{y}|$ for any null $j$.

Now imagine that we run the OMP or the Lasso on the original design augmented with knockoffs—that is, on the response $\mathbf{y}$ and the design matrix $\begin{bmatrix}\mathbf{X} & \widetilde{\mathbf{X}}\end{bmatrix}$, rather than on the smaller matrix $\mathbf{X}$. Let $[\hat{\boldsymbol{\beta}}\ \tilde{\boldsymbol{\beta}}]\in\mathbb{R}^{2p}$ denote the resulting coefficient estimates. Due to (6), we can see that $\hat{\beta}_j$ and $\tilde{\beta}_j$ have identical distributions whenever $j$ is null, and therefore, a false positive on feature $j$ (i.e. $\hat{\beta}_j\neq 0$) is equally likely to selecting the knockoff of feature $j$ (i.e. $\tilde{\beta}_j\neq 0$). Therefore, the number of knockoff variables that are in our estimated coefficient vector provides an estimate of the number of false positives.

## 2.2 The knockoff filter

After constructing knockoff copies of the features, we then use them to create a feature selection rule that will achieve exact FDR control. (One can also control other types of error such as the family-wise error rate (FWER) or the $k$-FWER as in [15]). For each variable $j\in\{1,\ldots,p\}$, we work with a general statistic $W_j$ obeying a *sufficiency* and an *antisymmetry* property. The sufficiency property requires that $W_j$ depends upon the design matrix only through its covariance, and upon the response $\mathbf{y}$ through marginal correlations: for some function $w_j$,

$$W_j = w_j(\begin{bmatrix}\mathbf{X} & \widetilde{\mathbf{X}}\end{bmatrix}^\top\begin{bmatrix}\mathbf{X} & \widetilde{\mathbf{X}}\end{bmatrix}, \begin{bmatrix}\mathbf{X} & \widetilde{\mathbf{X}}\end{bmatrix}^\top\mathbf{y}). \tag{7}$$

The antisymmetry property requires that swapping $\mathbf{X}_j$ and $\widetilde{\mathbf{X}}_j$ has the effect of changing the sign of $W_j$,

$$w_j(\begin{bmatrix}\mathbf{X} & \widetilde{\mathbf{X}}\end{bmatrix}_{\text{swap}(S)},\mathbf{y}) = \begin{cases}w_j(\begin{bmatrix}\mathbf{X} & \widetilde{\mathbf{X}}\end{bmatrix},\mathbf{y}), & j\notin S, \\ -w_j(\begin{bmatrix}\mathbf{X} & \widetilde{\mathbf{X}}\end{bmatrix},\mathbf{y}), & j\in S.\end{cases} \tag{8}$$

---

[3]If $\mathbf{s}$ is chosen so that the Gram matrix is positive semidefinite, then $\widetilde{\mathbf{X}}$ can be constructed to satisfy this matrix equality as long as $n\geq 2p$.

In [1], a much discussed example of such a statistic concerns knots at which variables enter the Lasso path: for each feature $\mathbf{X}_j$ and each knockoff $\widetilde{\mathbf{X}}_j$, record the first time that this feature or its knockoff enters the Lasso path, i.e. the largest penalty parameter value $\lambda$ such that $\hat{\beta}_j \neq 0$ or $\tilde{\beta}_j \neq 0$; then set

$$W_j = \left(\text{largest } \lambda \text{ such that } \mathbf{X}_j \text{ or } \widetilde{\mathbf{X}}_j \text{ enters Lasso path}\right) \cdot \begin{cases} +1, & \text{if } \mathbf{X}_j \text{ enters before } \widetilde{\mathbf{X}}_j, \\ -1, & \text{if } \mathbf{X}_j \text{ enters after } \widetilde{\mathbf{X}}_j. \end{cases} \tag{9}$$

We can analogously record the first time that a variable $\mathbf{X}_j$ or $\widetilde{\mathbf{X}}_j$ is selected by the OMP method or another forward stepwise algorithm, or any penalized least-squares method (replacing the $\ell_1$ penalty of the lasso with another penalty, such as $\ell_q$). Many other statistics are of course possible, for instance $W_j = |\hat{\beta}_j| - |\tilde{\beta}_j|$.

In all of these examples, a large positive $W_j$ indicates that $\mathbf{X}_j$ was strongly preferred over $\widetilde{\mathbf{X}}_j$ by the algorithm, while a large negative $W_j$ indicates $\widetilde{\mathbf{X}}_j$ was preferred. The sufficiency and antisymmetry properties of $W$, together with the exchangeability property (6) of the variables and knockoffs, ensures that $W_j > 0$ and $W_j < 0$ are equally likely for each null $j$, and in fact,

$$\#\{\text{null } j : W_j \leq -t\} \stackrel{d}{=} \#\{\text{null } j : W_j \geq t\}. \tag{10}$$

We use this property to choose a threshold for the test statistics $W_j$. Define a threshold $T > 0$ by setting[4]

$$T = \min\left\{t > 0 : \frac{\#\{j : W_j \leq -t\}}{\#\{j : W_j \geq t\}} \leq q\right\}, \tag{11}$$

where $q$ is the target FDR level. By (10), the numerator in (11) is an (over)estimate for the number of false discoveries among all reported variables with $W_j \geq t$, and so the ratio appearing in (11) is an (over)estimate of the false discovery proportion (FDP) at the threshold $t$. The output of the procedure the selected model

$$\widehat{S} = \{j : W_j \geq T\}, \tag{12}$$

which has FDP estimated to be $\leq q$. A slightly more conservative procedure, the knockoff+ filter, instead uses the threshold

$$T_+ = \min\left\{t > 0 : \frac{1 + \#\{j : W_j \leq -t\}}{\#\{j : W_j \geq t\}} \leq q\right\}, \tag{13}$$

and accordingly sets

$$\widehat{S} = \{j : W_j \geq T_+\}. \tag{14}$$

The key results in [1] state that knockoff+ controls the FDR, and the knockoff controls a slight modification of the FDR, at the level $q$. These results hold without any assumptions on the design or the value of the unknown regression coefficients. Crucially, we emphasize that neither the knockoff procedure nor its FDR controlling property assume any knowledge of the noise level $\sigma$.

## 3 Controlling Errors of Type S

Although they are more difficult to control, we have argued that Type S errors are more meaningful for regression problems since we would like to be able to tell the direction of effect with confidence (and should avoid reporting those variables whose direction we cannot reliably determine). Consider the setting of Section 2 where $n \geq 2p$ and $\mathbf{y} \sim \mathcal{N}(\mathbf{X}\boldsymbol{\beta}, \sigma^2 \mathbf{I})$. Then to estimate the direction of effect—the sign of $\beta_j$—note that

$$(\mathbf{X}_j - \widetilde{\mathbf{X}}_j)^\top \mathbf{y} \stackrel{\perp}{\sim} \mathcal{N}(s_j \beta_j, 2 s_j \sigma^2) \text{ for } j = 1, \ldots, p,$$

where $\mathbf{s} \geq \mathbf{0}$ is from the knockoff construction (5). Hence, a natural estimate for the sign of $\beta_j$ is

$$\widehat{\text{sign}}_j = \text{sign}\left((\mathbf{X}_j - \widetilde{\mathbf{X}}_j)^\top \mathbf{y}\right). \tag{15}$$

---

[4]More formally, we minimize over $t \in \mathcal{W}_+$, where $\mathcal{W}_+ = \{|W_j| : |W_j| > 0, j = 1, \ldots, p\}$, to avoid ever choosing $T = 0$.

With this, the knockoff filter with no adjustment whatsoever controls the directional FDR as well as the FDR. To understand intuitively why this error measure, which in principle is harder to control than FDR, is still bounded by the same method, note that for a nonnegative effect ($\beta_j \geq 0$), the probability of choosing a negative sign, i.e. $\widehat{\text{sign}}_j = -1$, is in fact highest when $\beta_j = 0$; this error is less likely if $\beta_j > 0$ since in that case we would have $(\mathbf{X}_j - \widetilde{\mathbf{X}}_j)^\top \mathbf{y}$ be normally distributed with a positive mean.

**Theorem 1.** *Assume* $\mathbf{y} \sim \mathcal{N}(\mathbf{X}\boldsymbol{\beta}, \sigma^2 \mathbf{I})$ *and fix any desired FDR level* $q$. *With the estimated direction effects* (15), *knockoff+ controls the directional FDR* (2); *that is,* $\text{FDR}_{\text{dir}} \leq q$ *for the selected set* $\widehat{S}$ *defined in* (14). *If we use knockoff instead, then* $\text{mFDR}_{\text{dir}} \leq q$, *where* $\text{mFDR}_{\text{dir}}$ *is the slightly modified version defined as*

$$\text{mFDR}_{\text{dir}} = \mathbb{E}\left[\frac{\left|\left\{j \in \widehat{S} : \widehat{\text{sign}}_j \neq \text{sign}(\beta_j)\right\}\right|}{|\widehat{S}| + q^{-1}}\right], \quad (16)$$

*where the selected set* $\widehat{S}$ *is defined as in* (12).

This result (proved in Appendix A.1) is more powerful than Theorems 1 and 2 from [1], which established that knockoff+ controls the FDR at level $q$, i.e. $\text{FDR} \leq q$, and knockoff controls the modified FDR, i.e. the expected ratio between the number of (unsigned) false discoveries $|\{j \in \widehat{S} \text{ and } \beta_j = 0\}|$ and $|\widehat{S}| + q^{-1}$ (this denominator is as in (16)).

Our new Theorem 1 is also more delicate to prove. The reason is that when we work with Type I errors as in the original result, the probability of claiming that $\beta_j \neq 0$ is calculated under the assumption that $\beta_j = 0$. When working with Type S errors, however, there is no such calibration since there is no null, and each feature $\mathbf{X}_j$ might have a nonzero coefficient $\beta_j$, although it may be close to zero. In that case, the test statistic $W_j$ will have a different probability of appearing with a positive or negative sign, and we are no longer dealing with i.i.d. and unbiased signs for the null $W_j$'s.

Thus in the proof of Theorem 1, we will see that the result is ultimately a consequence of the following martingale inequality regarding a sequence of Bernoulli variables which have varying probabilities of success:

**Lemma 1.** *Suppose that* $B_1, \ldots, B_n$ *are independent variables, with* $B_i \sim \text{Bernoulli}(\rho_i)$ *for each* $i$, *where* $\min_i \rho_i \geq \rho > 0$. *Let* $J$ *be a stopping time in reverse time with respect to the filtration* $\{\mathcal{F}_j\}$, *where*

$$\mathcal{F}_j = \{B_1 + \cdots + B_j, B_{j+1}, \ldots, B_n\}.$$

*Then*

$$\mathbb{E}\left[\frac{1+J}{1+B_1+\cdots+B_J}\right] \leq \rho^{-1}.$$

The proof of this lemma is given in Appendix B. In the simpler setting where $\rho_1 = \cdots = \rho_n$, this result is proved (in a slightly different form) in [1], with the proof relying heavily on the exchangeability of the $B_j$'s (when the $B_j$'s are determined by the signs of the null $W_j$'s, each has distribution Bernoulli(0.5)). The result stated here is more subtle due to the lack of exchangeability.

To see the connection between Lemma 1 and (directional) FDR control for the knockoff+ method, observe that

$$\text{FDP} = \frac{\#\{j \text{ null} : W_j \geq T_+\}}{1 \vee \#\{j : W_j \geq T_+\}} = \frac{\#\{j \text{ null} : W_j \geq T_+\}}{1 + \#\{j \text{ null} : W_j \leq -T_+\}} \cdot \frac{1 + \#\{j \text{ null} : W_j \leq -T_+\}}{1 \vee \#\{j : W_j \geq T_+\}}$$

$$\leq q \cdot \frac{\#\{j \text{ null} : W_j \geq T_+\}}{1 + \#\{j \text{ null} : W_j \leq -T_+\}},$$

where the inequality follows from the definition of $T_+$ (13). Now reorder the indices of the null $W_j$'s so that $|W_{(1)}| \geq |W_{(2)}| \geq \ldots$, where $|W_{(1)}|$ is the null with the largest magnitude, $|W_{(2)}|$ the second largest, and so on, and let $B_j = \mathbb{1}_{W_{(j)} < 0}$. Then we have

$$\frac{\#\{j \text{ null} : W_j \geq T_+\}}{1 + \#\{j \text{ null} : W_j \leq -T_+\}} = \frac{(1-B_1) + \cdots + (1-B_J)}{1 + B_1 + \cdots + B_J} = \frac{1+J}{1+B_1+\cdots+B_J} - 1,$$

where $J$ is the index such that $|W_{(1)}| \geq \cdots \geq |W_{(J)}| \geq T_+ > |W_{(J+1)}| \geq \ldots$; this last expression is the quantity in Lemma 1. If each $W_j$, for a null feature, is equally likely to be positive or negative, then the $B_j$'s are i.i.d. Bernoulli(0.5) variables, and we obtain FDR control at level $q$ by applying Lemma 1 with $\rho_i = 0.5$. However, when considering directional FDR control, the $\beta_j$'s may be nonzerzo and so the $W_j$'s will typically not be symmetric; therefore we may need to apply Lemma 1 with varying $\rho_i$'s. We refer to Appendix B.1 for details.

# 4 Knockoffs in High Dimensions

In high dimensions, where $p > n$, the knockoff construction is no longer possible—in fact, the Gram matrix condition (5) would be possible only if $\mathbf{s} = \mathbf{0}$, that is, if $\widetilde{\mathbf{X}}_j = \mathbf{X}_j$ for each feature $j = 1, \ldots, p$, so that the knockoff procedure would have zero power. In this setting, one straightforward approach would be to use part of the data to reduce the number of features, and a disjoint part of the data to run the knockoff filter. In the next section, we develop this data splitting approach, and then find that we can gain substantial power by using a subtle way of 'recycling' the data.

## 4.1 Feature screening and reusing data

Consider splitting the $n$ observations into two disjoint groups of size $n_0$ and $n_1 = n - n_0$, used for the purpose of first screening for a smaller set of potentially relevant features, then running a model selection procedure over this reduced list of features, respectively. We denote the two disjoint portions of the data as $(\mathbf{X}^{(0)}, \mathbf{y}^{(0)}) \in \mathbb{R}^{n_0 \times p} \times \mathbb{R}^{n_0}$ and $(\mathbf{X}^{(1)}, \mathbf{y}^{(1)}) \in \mathbb{R}^{n_1 \times p} \times \mathbb{R}^{n_1}$, and then follow the next two steps:

- **Screening step:** using $(\mathbf{X}^{(0)}, \mathbf{y}^{(0)})$, we identify a subset $\widehat{S}_0 \subset [p]$ of potentially relevant features, such that $|\widehat{S}_0| < n_1$.

- **Selection step (splitting):** ignoring any features that were discarded in the screening step, we run the knockoff procedure on the remaining data, that is, on $(\mathbf{X}^{(1)}_{\widehat{S}_0}, \mathbf{y}^{(1)})$.

This straightforward data splitting approach is a natural extension of the low-dimensional knockoff filter, and it is clear that the approach will control the directional FDR in the final model selection step as long as the screening step correctly captures all the relevant features—those with non-vanishing regression coefficients—a property often referred to as *sure screening* in the literature [9]. (False positives in the screening step, i.e. including too many null features, do not pose a problem). If the sure screening property fails, we can still obtain inference guarantees relative to the new, screened submodel—see Section 4.3 below.

There is an inherent loss of power due to the split of the data, since the model selection step uses $n_1$ rather than $n$ observations. However, using disjoint data sets for the screening and selection steps is critical, since the distribution of $\mathbf{y}^{(0)}$ cannot be treated as a Gaussian linear model *after* the screening step has taken place. The screening step is a function of the random variable $\mathbf{y}^{(0)}$, and so the following modification of the selection step *would not control the FDR*: ignoring any features that were discarded in the screening step, we run the knockoff procedure on the full data set, i.e. on $(\mathbf{X}_{\widehat{S}_0}, \mathbf{y})$. The loss of FDR control is not merely theoretical: a null feature $\mathbf{X}_j$ that is chosen by the screening step is generally more likely to appear as a false positive when running the knockoff filter, leading to a much higher FDR.

In light of this, we propose two mechanisms for increasing power relative to the data splitting procedure described so far:

1. *data recycling*, where the first portion of the split data can be reused to some extent without losing *any* of the guaranteed FDR control,

2. and *signed statistics*, where we can decide adaptively to focus our search on a positive effect only or a negative effect only.

#### 4.1.1 Increasing power with data recycling

Surprisingly, data recycling retains the FDR control properties of the split-data procedure, while raising power substantially to approach the sensitivity of the full-data procedure.

- **Selection step (with recycling):** we begin by constructing a knockoff matrix on the remaining data $\widetilde{\mathbf{X}}_{\widehat{S}_0}^{(1)}$ for $\mathbf{X}_{\widehat{S}_0}^{(1)}$ just as we would in the data splitting version. The difference now is that we concatenate the original design matrix on the first $n_0$ observations with the knockoff matrix on the next $n_1$ observations,

$$\widetilde{\mathbf{X}}_{\widehat{S}_0} = \begin{bmatrix} \mathbf{X}_{\widehat{S}_0}^{(0)} \\ \widetilde{\mathbf{X}}_{\widehat{S}_0}^{(1)} \end{bmatrix}. \tag{17}$$

(Note that on the first part of the split data, knockoffs are exact copies.) We then run the knockoff filter on the entire data set with all $n$ samples, using the design matrix $\mathbf{X}_{\widehat{S}_0}$, the knockoff matrix $\widetilde{\mathbf{X}}_{\widehat{S}_0}$, and the original response $\mathbf{y}$.

The term "recycling" refers to the way that we incorporate the data $(\mathbf{X}_{\widehat{S}_0}^{(0)}, \mathbf{y}^{(0)})$, which was already used in the screening step, for the selection step. Here, we think of the mechanism for choosing $\widehat{S}_0$ as being completely arbitrary as opposed to having some kind of 'pre-registered' screening method. For this reason, we need to treat $\mathbf{y}^{(0)}$ as a fixed vector (or rather, to condition on its value) in the construction and analysis of the knockoff filter for the selection step (see Section 4.1.3 for additional discussion about this point).

Of course, knockoffs with data splitting are actually a special case of knockoffs with data recycling. Any knockoff statistics $W_j$ that are a function of $\mathbf{X}_{\widehat{S}_0}^{(1)}$, $\widetilde{\mathbf{X}}_{\widehat{S}_0}^{(1)}$, and $\mathbf{y}^{(1)}$ (i.e. data splitting) can trivially be expressed as a function of $\mathbf{X}_{\widehat{S}_0}$, $\widetilde{\mathbf{X}}_{\widehat{S}_0}$, and $\mathbf{y}$ by simply ignoring the first $n_0$ many data points.

**Increased power** Why do we expect the knockoff method to exhibit higher power when using data recycling rather than data splitting, when we see in (17) that the knockoff matrix differs from the original features only on the second part of the split data? Indeed, later on in Section 5 we will see empirically that the power gain can be quite substantial.

To understand this phenomenon, we return to the mechanism of the knockoff filter itself. Recall that we construct a statistic $W_j$ for each feature/knockoff pair $(\mathbf{X}_j, \widetilde{\mathbf{X}}_j)$, where $|W_j|$ is large if *either* $\mathbf{X}_j$ or $\widetilde{\mathbf{X}}_j$ appear highly significant, and $\text{sign}(W_j)$ indicates which of the two appears most significant. For instance, if we are using a forward stepwise selection method as described in Section 2, $|W_j|$ will be large if either $\mathbf{X}_j$ or $\widetilde{\mathbf{X}}_j$ was one of the first variables to enter the model; we will have $W_j > 0$ if $\mathbf{X}_j$ enters before $\widetilde{\mathbf{X}}_j$, and $W_j < 0$ if $\mathbf{X}_j$ enters after $\widetilde{\mathbf{X}}_j$.

If the early portion of the knockoff path (where we order the features in order of magnitudes $|W_j|$) has many positive $W_j$'s and few negative $W_j$'s, the stopping rule (11) will allow us to choose a threshold $T$ that is not too large, and we will make many rejections. If instead the high-magnitude $|W_j|$'s have many negative signs, though, we will be forced to stop early, choosing a high threshold $T$ and making few rejections. For any null feature $j$ (i.e. $\beta_j = 0$), the sign of $W_j$ is equally likely to be negative as positive, so in order to obtain high power, we need two things:

1. Good separation in the feature ordering—most of the features that appear early in the path, i.e. most $j$ with $|W_j|$ large, are non-nulls;
2. Good power in the non-null signs—for most non-nulls, the statistics have a positive sign, $W_j > 0$.

Next, we ask how data splitting and data recycling compare in light of these two considerations. For the second, both methods suffer from power loss relative to an unscreened method—specifically, obtaining $W_j > 0$ rather than $W_j < 0$ depends on the model selection method being able to distinguish between feature $j$ and its knockoff copy. For data splitting, by the sufficiency (7) and antisymmetry (8) properties, we see that if

$$\mathbf{X}_j^{(1)\top}\mathbf{y}^{(1)} \quad \text{and} \quad \widetilde{\mathbf{X}}_j^{(1)\top}\mathbf{y}^{(1)} \tag{18}$$

are equal in distribution, then $\text{sign}(W_j)$ is equally likely to be $+1$ or $-1$; in other words, it is only if these two inner products have substantially *different* distributions, that we can hope to see $\text{sign}(W_j) = +1$ with high probability. For data recycling, the same statement holds, except instead we compare the inner products

$$\mathbf{X}_j^\top \mathbf{y} = \mathbf{X}_j^{(0)\top} \mathbf{y}^{(0)} + \mathbf{X}_j^{(1)\top} \mathbf{y}^{(1)} \quad \text{and} \quad \widetilde{\mathbf{X}}_j^\top \mathbf{y} = \mathbf{X}_j^{(0)\top} \mathbf{y}^{(0)} + \widetilde{\mathbf{X}}_j^{(1)\top} \mathbf{y}^{(1)}. \tag{19}$$

Examining (18) versus (19), we see that these two comparisons are equivalent, as (19) simply adds the same constant (i.e. $\mathbf{X}_j^{(0)\top} \mathbf{y}^{(0)}$) to each term of (18). In other words, these two tasks are equally difficult.

However, for the first component of power—that is, good separation in the feature ordering—we can expect that knockoffs with data recycling will have the advantage, since the first $n_0$ data points carry substantial information separating signals from nulls. Of course, if some non-null feature $j$ has strong correlation with the response, then the information from the first $n_0$ data points will push *both* $\mathbf{X}_j$ and $\widetilde{\mathbf{X}}_j$ towards the top of the path—since $\widetilde{\mathbf{X}}_j^{(0)} = \mathbf{X}_j^{(0)}$ by construction—but the second portion of the data, where $\widetilde{\mathbf{X}}_j^{(1)} \neq \mathbf{X}_j^{(1)}$, will hopefully enable $\mathbf{X}_j$ to enter the path before $\widetilde{\mathbf{X}}_j$ (i.e. $W_j > 0$).

To summarize, the knockoff filter's power relies on the non-null $W_j$'s ability to *appear early* and *with positive sign* along the path of statistics; data recycling helps the non-null $W_j$'s appear early, thus increasing power.

### 4.1.2 Increasing power with signed statistics

Since the knockoff filter with data splitting or with data recycling treats the first portion of the data, $(\mathbf{X}^{(0)}, \mathbf{y}^{(0)})$, as fixed once the screening step has been performed, we are free to use this data in any way we wish, to try to gather information about the true signals. While the screening step identifies the indices of the variables which are likely to contain signals, we can also use this step to identify probable signs—that is, if feature $\mathbf{X}_j^{(0)}$ is selected in the screening step, which is run on the data set $(\mathbf{X}^{(0)}, \mathbf{y}^{(0)})$, we can at the same time specify the sign of the estimated effect of this feature, denoted $\widehat{\text{sign}}_j^{(0)}$. We then use this information to look specifically for an effect consistent with this sign in the second phase of our procedure. This will in general increase power, since we are extracting more information from the first part of the data before running the selection procedure on the second part, $(\mathbf{X}^{(1)}, \mathbf{y}^{(1)})$.

**Example: penalized least-squares** To illustrate the use of the sign information in practice, suppose that we are using a penalized least-squares regression method for computing the statistics $W_j$—for instance, as described in Section 2, the $W_j$'s may indicate the time at which features $\mathbf{X}_j$ and $\widetilde{\mathbf{X}}_j$ entered the Lasso path (or the path of some other sequential selection procedure), or may be given by $W_j = |\hat{\beta}_j| - |\tilde{\beta}_j|$ where $[\hat{\boldsymbol{\beta}} \ \tilde{\boldsymbol{\beta}}]$ is the solution to a penalized least squares optimization problem, with some penalty function $P(\mathbf{b})$ (with the Lasso as a special case if $P(\mathbf{b}) = \lambda \|\mathbf{b}\|_1$). In either case, if we are not using sign information, we might begin by computing the penalized least-squares solution $[\hat{\boldsymbol{\beta}} \ \tilde{\boldsymbol{\beta}}]$ via either

$$\min_{\mathbf{b} \in \mathbb{R}^{2|\widehat{S}_0|}} \quad \tfrac{1}{2} \|\mathbf{y}^{(1)} - [\mathbf{X}_{\widehat{S}_0}^{(1)} \ \widetilde{\mathbf{X}}_{\widehat{S}_0}^{(1)}] \mathbf{b}\|_2^2 + P(\mathbf{b})$$

for data splitting, or

$$\min_{\mathbf{b} \in \mathbb{R}^{2|\widehat{S}_0|}} \quad \tfrac{1}{2} \|\mathbf{y} - [\mathbf{X}_{\widehat{S}_0} \ \widetilde{\mathbf{X}}_{\widehat{S}_0}] \mathbf{b}\|_2^2 + P(\mathbf{b})$$

for data recycling. Instead, to make use of the sign information gathered at the screening phase, we can consider a sign-restricted version of this penalized least-squares problem: in the data splitting version we would be interested in the solution to

$$\begin{aligned}
\min_{\mathbf{b} \in \mathbb{R}^{2|\widehat{S}_0|}} \quad & \tfrac{1}{2} \|\mathbf{y}^{(1)} - [\mathbf{X}_{\widehat{S}_0}^{(1)} \ \widetilde{\mathbf{X}}_{\widehat{S}_0}^{(1)}] \mathbf{b}\|_2^2 + P(\mathbf{b}) \\
\text{s.t.} \quad & b_j \cdot \text{sign}_j^{(0)} \geq 0 \\
& b_{j+|\widehat{S}_0|} \cdot \text{sign}_j^{(0)} \geq 0
\end{aligned}$$

and similarly for the recycling version. In other words, we are running the same penalized least-squares optimization, but with the added restriction that we will only select the $j$th feature or $j$th knockoff feature if its estimated effect direction agrees with the sign information gathered at the screening stage.

**Increased power** Why should we expect the knockoff method to exhibit higher power when leveraging the sign information, $\text{sign}_j^{(0)}$ for each selected feature $j \in \widehat{S}_0$? Intuitively, we think of the screening step as choosing from $p$ possible hypotheses, where hypothesis $H_j$ is described by the question, "Does $\mathbf{X}_j$ appear in the true model?". However, we can instead frame this step as considering twice as many hypotheses: hypothesis $H_j^+$ is the question, "Does $\mathbf{X}_j$ appear in the true model with a positive effect?", and hypothesis $H_j^-$ is the question "Does $\mathbf{X}_j$ appear in the true model with a negative effect?". With this in mind, if the screening step fits a model to the first part of the data $(\mathbf{X}^{(0)}, \mathbf{y}^{(0)})$, and this fitted model includes the $j$th feature with a positive coefficient, then this indicates that hypothesis $H_j^+$ is more likely to contain a signal. Therefore, it makes sense that in our selection step we should focus our attention on hypothesis $H_j^+$. Of course, it is possible that the true effect is in fact negative, but since this is less likely, we would generally be increasing the noise if we give equal attention to hypotheses $H_j^+$ and $H_j^-$.

#### 4.1.3 Relation to existing work

We pause here to compare our data recycling technique to the data carving methods of Fithian et al. [10]. In that work, as in ours, some part of the data $(\mathbf{X}^{(0)}, \mathbf{y}^{(0)})$ is used to perform an initial screening step, and is then reused for inference along with the remaining data $(\mathbf{X}^{(1)}, \mathbf{y}^{(1)})$. However, the mechanism behind reusing the first part of the data is highly different. In Fithian et al. [10]'s work, the first part of the response $\mathbf{y}^{(0)}$ is reused by leveraging the "remaining randomness" after the screening step: $\mathbf{y}^{(0)}$ is regarded as a random vector, with distribution given by conditioning on the outcome of the screening step, i.e. $(\mathbf{y}^{(0)} \mid \widehat{S}_0)$. In contrast, in our "recycling" procedure, $\mathbf{y}^{(0)}$ can be used in an arbitrary way for the screening step, and so there is no "remaining randomness" in $\mathbf{y}^{(0)}$. Instead, we reuse $\mathbf{y}^{(0)}$ by treating it as a fixed vector; $\mathbf{y}^{(1)}$ is the only variable treated as random for the purpose of our FDR control results.

More broadly, the knockoff filter with either data splitting or data recycling, can be considered as similar in flavor to the "screen and clean" methodology of Wasserman and Roeder [35], where an initial screening step (performed via a high-dimensional Lasso, with a tuning parameter chosen by validation), is followed by an inference step on an independent portion of the data, with inference performed via a least-squares regression. This methodology has been applied to genome-wide association studies (GWAS) by Wu et al. [37]. At a high level, our method follows this same overall framework of screening for a low-dimensional submodel, then using new data for low-dimensional inference; however, we will gain power by "recycling" the first part of the data, and more significantly, by using the knockoff filter rather than least-squares for the inference step, which in the low-dimensional setting gives substantial gains in power as it leverages the sparse structure of the true model [1].

### 4.2 Sure screening and directional FDR control

In order to demonstrate the correctness of the selection step with recycling, we first assume that the screening step is highly likely to find all true signals—that is, $\widehat{S}_0 \supseteq \text{support}(\boldsymbol{\beta})$ with high probability. Note that we do not assume that the screening step *exactly* selects the true support—while this type of sparsistency result holds for the Lasso in the asymptotic setting (see e.g. [41]), in practice, even with ideal simulated data, it is generally the case that we cannot perfectly separate true signals from false positives using the limited data available, unless the signal is extremely strong and extremely sparse [28]. Instead, this assumption merely requires that there is some sufficiently liberal screening procedure that is likely to capture all the true signals, along with many false positives. From a theoretical standpoint, this property is known to hold under conditions far weaker than those needed for exact recovery of the true support [39, 9].

Suppose we fix some chosen method for the screening step, which is constrained only in that it must be a function of the first portion of the data, $(\mathbf{X}^{(0)}, \mathbf{y}^{(0)})$. Define the sure screening event as

$$\mathcal{E} = \{\widehat{S}_0 \supseteq \text{support}(\boldsymbol{\beta}) \text{ and } |\widehat{S}_0| \leq n_1/2\},$$

and note that $\mathbb{1}_\mathcal{E}$ is a function of $\mathbf{y}^{(0)}$ when we treat $\mathbf{X}$ and $\boldsymbol{\beta}$ as fixed. Conditioning on this event, directional FDR control of the selection step holds by the following theorem (proved in Appendix A.1):

**Theorem 2.** *Suppose* $\mathbf{y} \sim \mathcal{N}(\mathbf{X}\boldsymbol{\beta}, \sigma^2 \mathbf{I})$. *Then the knockoff procedure (using either data splitting or data recycling), with estimated signs* $\widehat{\text{sign}}_j$ *as in* (15), *controls the modified directional FDR at the level*

$$\mathbb{E}\left[\text{mFDR}_{\text{dir}} \mid \mathcal{E}\right] \leq q \,,$$

*while if knockoff+ is used, then the directional FDR is controlled as*

$$\mathbb{E}\left[\text{FDR}_{\text{dir}} \mid \mathcal{E}\right] \leq q \,.$$

In particular, if $\mathcal{E}$ occurs with probability near 1, then the various forms of the directional FDR are controlled even without conditioning on $\mathcal{E}$, by reformulating the results as $\text{FDR}_{\text{dir}} \leq q + \mathbb{P}\{\mathcal{E}^c\}$ in the case of knockoff+ and $\text{mFDR}_{\text{dir}} \leq q + \mathbb{P}\{\mathcal{E}^c\}$ for knockoff.

## 4.3 Directional FDR control in the reduced model

If the screening step misses variables from the model, a bias is introduced which creates a fundamental difficulty. To discuss this, recall that knockoffs are constructed to guarantee the crucial pairwise exchangeability for the nulls (6). Among other things, we have seen that exchangeability has the consequence that null variables are equally likely to be selected as their knockoff companions. When nonzero effects have not been screened, this may no longer be the case. Concretely, imagine that $\mathbf{X}_k$ has a nonzero effect but was not identified in the screening step. Then since the knockoff construction used only the screened set $\widehat{S}_0$, we cannot guarantee that $\mathbf{X}_j^\top \mathbf{X}_k = \widetilde{\mathbf{X}}_j^\top \mathbf{X}_k$ for $j \in \widehat{S}_0$, which in turn may lead to

$$\mathbf{X}_j^\top \mathbf{y} \stackrel{d}{\neq} \widetilde{\mathbf{X}}_j^\top \mathbf{y}$$

even when $j$ is a null feature, i.e. $\beta_j = 0$. For example, if $\mathbf{X}_j$ were strongly correlated with a true signal $\mathbf{X}_k$, then we could easily imagine that $\mathbf{X}_j$ would be more likely to be selected than its knockoff $\widetilde{\mathbf{X}}_j$. In general, and in contrast to the lower dimensional setting, the bias implies that when we count the number of knockoffs selected at some threshold level, we cannot be sure that this number is a good (over)estimate of the number of selected nulls. (As explained before, however, it may be desirable to select $\mathbf{X}_j$ in this case since it is a proxy for $\mathbf{X}_k$.)

We now reframe this issue in terms of the partial regression coefficients—that is, we will perform inference on the coefficients in the reduced model defined by the subset of features $\widehat{S}_0$ selected in the screening step, a point of view in line with much of the recent literature discussed earlier [20, 7, 12, 10]. Thus we are interested in the partial regression coefficients—the coefficients of $\mathbf{y}$ regressed onto only the selected features. Throughout this section, we will consider a random design model, and will define the population-level partial regression coefficients as

$$\boldsymbol{\beta}^{\text{partial}} = (\mathbf{X}_{\widehat{S}_0}^{(1)\top} \mathbf{X}_{\widehat{S}_0}^{(1)})^{-1} \mathbf{X}_{\widehat{S}_0}^{(1)\top} \mathbb{E}\left[\mathbf{y}^{(1)} \mid \mathbf{X}_{\widehat{S}_0}^{(1)}; \mathbf{X}^{(0)}, \mathbf{y}^{(0)}\right], \tag{20}$$

that is, the *expected* coefficients when $\mathbf{y}^{(1)}$ regressed on the selected features $\mathbf{X}_{\widehat{S}_0}^{(1)}$. In other words, this is the quantity for which the least-squares solution in the reduced model is an unbiased estimate. Here, expectation is taken conditional on the first part of the data (i.e. $(\mathbf{X}^{(0)}, \mathbf{y}^{(0)})$, the data used for the screening step), and on the selected features $\mathbf{X}_{\widehat{S}_0}^{(1)}$.[5]

**FDR control with a Gaussian design** In this paper, we give results for a random feature model $\mathbf{X}$, which assumes that the rows $\mathbf{X}_{[i]}$ of $\mathbf{X}$ are drawn i.i.d. from a multivariate Gaussian distribution,

$$\mathbf{X}_{[i]} \stackrel{\text{iid}}{\sim} \mathcal{N}(\boldsymbol{\nu}, \boldsymbol{\Psi}); \tag{21}$$

---

[5]This definition is slightly different from the notion of partial regression coefficients found in much of the selective inference literature discussed here. i.e. [20, 7, 12, 10], which treats the design $\mathbf{X}$ as fixed and studies $\boldsymbol{\beta}^{\text{partial}} = (\mathbf{X}_{\widehat{S}_0}^\top \mathbf{X}_{\widehat{S}_0})^{-1} \mathbf{X}_{\widehat{S}_0}^\top \mathbb{E}[\mathbf{y} \mid \mathbf{X}]$. We could alternately consider a population-level version of these coefficients, $\boldsymbol{\beta}^{\text{partial}} = \mathbb{E}\left[\mathbf{X}_{\widehat{S}_0}^\top \mathbf{X}_{\widehat{S}_0}\right]^{-1} \mathbb{E}\left[\mathbf{X}_{\widehat{S}_0}^\top \mathbf{y}\right]$. The difference is between these various definitions is typically not substantial, and indeed under many random models for the design $\mathbf{X}$, can be proved to be vanishingly small.

here, the parameters (i.e. the mean vector $\boldsymbol{\nu} \in \mathbb{R}^p$ and the covariance matrix $\boldsymbol{\Psi} \in \mathbb{R}^{p \times p}$) are arbitrary and completely unknown to us. The following theorem (proved in Appendix A.1) guarantees directed FDR control in this setting.

**Theorem 3.** *Assume that the rows of $\mathbf{X}$ follow* (21)*, where $\boldsymbol{\nu} \in \mathbb{R}^p$ and $\boldsymbol{\Psi} \in \mathbb{R}^{p \times p}$ are arbitrary and unknown. Define the expected partial regression coefficients $\boldsymbol{\beta}^{\text{partial}}$ as in* (20)*. Then the knockoff-with-recycling procedure controls the modified directional FDR at the level*

$$\text{mFDR}_{\text{dir}} \leq q \;,$$

*while if knockoff+ is used, then the directional FDR is controlled as*

$$\text{FDR}_{\text{dir}} \leq q \;;$$

*here, the estimated signs $\widehat{\text{sign}}_j$ are defined as in* (15)*, and a false discovery is any $j \in \widehat{S}$ such that $\widehat{\text{sign}}_j \neq \text{sign}(\beta_j^{\text{partial}})$.*

While the Gaussian design assumption is strong, we believe that FDR control would be maintained across a far broader range of models.[6] We emphasize that we do not assume knowledge of the mean and covariance parameters of the Gaussian distribution of $\mathbf{X}$, and do not need to estimate these parameters in the method; in fact the Gaussian assumption is used only for the technical details of the proof, to show that we can model the response $\mathbf{y}^{(1)}$ as

$$\mathbf{y}^{(1)} = \mathbf{X}_{\widehat{S}_0}^{(1)} \boldsymbol{\beta}_{\widehat{S}_0} + \mathbf{X}_{\widehat{S}_0^c}^{(1)} \boldsymbol{\beta}_{\widehat{S}_0^c} + \boldsymbol{\epsilon}^{(1)} = \mathbf{X}_{\widehat{S}_0}^{(1)} \boldsymbol{\beta}_{\widehat{S}_0}^{\text{partial}} + \boldsymbol{\epsilon}'^{(1)} \;,$$

where $\boldsymbol{\epsilon}'^{(1)}$ is i.i.d. Gaussian noise (with a larger variance than the original noise $\boldsymbol{\epsilon}$, due to the noise added from the true signals missed by the screening step). This is where the fact that knockoffs do not need the value of the noise level (in low dimensions) is immensely useful: the knockoff method will provide valid inference no matter the value of the 'new noise level'.

# 5 Simulations

We now examine the performance of our method on simulated high-dimensional data sets, and compare to several other techniques for inference in low-dimensional and high-dimensional regression. One data set below uses real SNP features.

## 5.1 Data

To simulate data that exhibits challenging features that we might expect to see in practice, we work in two different settings for the matrix $\mathbf{X}$ of covariates:

- **AR matrix**  We consider an autoregressive model, where $\mathbf{X}_j$ is highly correlated with $\mathbf{X}_{j+1}$ and $\mathbf{X}_{j-1}$, and the correlation decays for columns farther apart in the design matrix. We work with sample size $n = 2000$ and create $p = 2500$ features by drawing the rows of $\mathbf{X}$ independently from a $\mathcal{N}(\mathbf{0}, \boldsymbol{\Sigma})$ model, where $\Sigma_{jk} = \rho^{|j-k|}$ for a correlation parameter $\rho \in \{0, 0.25, 0.5, 0.75\}$. We then normalize the columns of $\mathbf{X}$ to have unit norm.

- **GWAS matrix**  The matrix $\mathbf{X}$ is here taken from a genome-wide association study (GWAS), see Section 6 for further information regarding this data set. We begin with SNP data from the first three chromosomes, then cluster the SNPs using pairwise correlations and choose only one SNP from each cluster, in such a way that the resulting pairwise correlation between any two SNPs in the remaining data set is at most 0.5, and finally regress out the top five principal components (Section 6 discusses the rationale behind this extra step). The size of the data set is $p = 12752$ features (SNPs) and sample size $n = 4682$.

---

[6] An additional result appears in an earlier draft of this work, available at https://arxiv.org/abs/1602.03574v1 (see Theorem 3), treating the case where the design matrix $\mathbf{X}$ is fixed and sure screening is not assumed, but instead assuming that the noise level $\sigma$ is known.

15                                                    02/2016; Revised 09/2017 and 05/2018

We then generate the coefficient vector $\boldsymbol{\beta}$ by choosing $k_0 = 50$ many coordinates at random to be the locations of the strong signals, and $k_1$ many coordinates to be the location of the weak signals, where $k_1 = 250$ for the AR setting and $k_1 = 1250$ for the GWAS setting. Strong signals are chosen from $\{\pm 4.5\}$ for the AR setting or from $\{\pm 5\}$ for the GWAS setting (the higher amplitude is due to the larger size of the problem). Weak signals are drawn as $\beta_j \sim \mathcal{N}(0, 0.5)$ for both settings. The response is then generated as $\mathbf{y} = \mathbf{X}\boldsymbol{\beta} + \boldsymbol{\epsilon}$, where $\boldsymbol{\epsilon}$ has independent $\mathcal{N}(0,1)$ entries. In each setting (AR design with each value of $\rho$, and the GWAS design), the design matrix $\mathbf{X}$ and the coefficient vector $\boldsymbol{\beta}$ are generated only once, while the noise vector $\boldsymbol{\epsilon}$ is generated independently for each trial. Results are averaged over 100 trials. All procedures are run with target FDR level $q = 0.2$.

## 5.2 Methods

Our simulated experiments compare the following methods:

- The Benjamini-Hochberg (BH) [4] procedure (applied to the least-squares regression coefficients, after screening for a low-dimensional submodel).

- Exact selective inference for the square-root Lasso, developed by Tian et al. [30]—this method builds on the earlier work of Lee et al. [20] developing selective inference for the Lasso.

- Knockoff filter with data splitting or data recycling (applied after screening for a low-dimensional submodel).

We now give details for the implementation of each method.

**Benjamini-Hochberg**  We use $n_0$ many data points for an initial screening step, where $n_0 = 750$ for the AR setting and $n_0 = 1800$ for the GWAS setting. The screening step is carried out by solving the Lasso across the entire range of $\lambda$ values, from $\lambda = \infty$ to $\lambda = 0$. We then choose a screened set $\widehat{S}_0$ consisting of the first $k_{\max}$ many features entering the path, where $k_{\max} = 450$ for the AR setting and $k_{\max} = 720$ for the GWAS setting, and then record the sign $s_j \in \{\pm 1\}$ for feature $j \in \widehat{S}_0$ (i.e. the sign of its estimated coefficient, at the time when it first enters the Lasso path). We next use the remaining $n_1 = n - n_0$ data points for inference. We calculate the least-squares regression coefficients $\hat{\boldsymbol{\beta}}_{\mathsf{LS}} = (\mathbf{X}^{(1)\top}_{\widehat{S}_0} \mathbf{X}^{(1)}_{\widehat{S}_0})^{-1} \mathbf{X}^{(1)\top}_{\widehat{S}_0} \mathbf{y}^{(1)}$, and compute t-scores

$$t_j = \frac{(\hat{\boldsymbol{\beta}}_{\mathsf{LS}})_j}{\hat{\sigma} \cdot \sqrt{(\mathbf{X}^{(1)\top}_{\widehat{S}_0} \mathbf{X}^{(1)}_{\widehat{S}_0})^{-1}_{jj}}}$$

for each $j \in \widehat{S}_0$. Above, $\hat{\sigma}^2$ is the classical estimate of variance,

$$\hat{\sigma}^2 = \frac{\|\mathbf{y}^{(1)} - \mathbf{X}^{(1)}_{\widehat{S}_0} \hat{\boldsymbol{\beta}}_{\mathsf{LS}}\|_2^2}{n_1 - |\widehat{S}_0|}.$$

Under the sure screening property (where $\widehat{S}_0$ contains all true signals, as discussed in Section 4.2), for each null feature $j$, $t_j$ would follow a $t$-distribution with $n_1 - |\widehat{S}_0|$ degrees of freedom. We thus convert each t-score to a one-sided p-value $P_j$ (with the side chosen based on the sign $s_j$ from the screening step). Finally, we run the Benjamini-Hochberg procedure [4] at the level $q$ to select a model.

**Selective inference for the square-root Lasso**  The square-root Lasso [2] is given by

$$\hat{\boldsymbol{\beta}} = \operatorname*{arg\,min}_{\mathbf{b} \in \mathbb{R}^p} \{\|\mathbf{y} - \mathbf{X}\mathbf{b}\|_2 + \lambda \cdot \|\mathbf{b}\|_1\} \text{ where } \lambda = \kappa \cdot \mathbb{E}_{\mathbf{g} \sim \mathcal{N}(0, \mathbf{I}_n)} \left[\frac{\|\mathbf{X}^\top \mathbf{g}\|_\infty}{\|\mathbf{g}\|_2}\right] \quad (22)$$

for a constant $\kappa$ typically chosen in $[0.5, 1]$. The solution $\hat{\boldsymbol{\beta}}$ is in fact *exactly* equal to the solution to the Lasso (4) at some different penalty parameter value; the benefit of the square-root Lasso is that the penalty parameter $\lambda$ can be chosen without reference to the unknown noise level $\sigma^2$.

Tian et al. [30] derive exact post-selection confidence intervals and p-values for the coefficients $\beta_j$, after conditioning on the selection event—that is, after observing the signed support of $\hat{\boldsymbol{\beta}}$, the solution to the square-root Lasso—based on the truncated $t$ distribution, without knowledge of the noise level $\sigma$. (This work builds on the polytope method developed in Lee et al. [20], Tibshirani et al. [32] for the Lasso and other procedures.) An implementation of this method, in Python, is available in [29], where the inference is performed approximately by estimating the noise $\sigma$ and then using truncated normal distributions, as in Lee et al. [20]'s inference for the Lasso. We use this available code for our experiments, in order to calculate a one-sided selective p-value $P_j$ for each feature selected by the square-root Lasso (based on the sign $s_j$ with which feature $j$ was selected). We then apply the Benjamini-Hochberg procedure at level $q$ to the resulting set of p-values in order to select a final model. The only tuning parameter in this method is the constant $\kappa$ appearing in (22). We test $\kappa \in \{0.5, 0.6, \ldots, 1\}$ and find the highest power at $\kappa = 0.6$; only this value of $\kappa$ is displayed in our results.[7]

**Knockoff filter**  The screening step is carried out exactly as for the BH procedure, with the same values of $n_0$ and $k_{\max}$, resulting in a submodel $\widehat{S}_0$ and signs $s_j$ for each $j \in \widehat{S}_0$. We next use the remaining $n_1 = n - n_0$ data points for the knockoff filter, using sign information as described in Section 4.1.2, and either data splitting or data recycling. Our statistics $W_j$ are formed using the square-root Lasso. Specifically, for data splitting, we solve a sign-restricted square-root Lasso,

$$[\hat{\boldsymbol{\beta}} \ \tilde{\boldsymbol{\beta}}] = \underset{\substack{\mathbf{b} \in \mathbb{R}^{2|\widehat{S}_0|} \\ b_j \cdot s_j \geq 0 \text{ for all } j=1,\ldots,p \\ b_{j+|\widehat{S}_0|} \cdot s_j \geq 0 \text{ for all } j=1,\ldots,p}}{\arg\min} \left\{ \|\mathbf{y}^{(1)} - [\mathbf{X}^{(1)}_{\widehat{S}_0} \ \widetilde{\mathbf{X}}^{(1)}_{\widehat{S}_0}]\mathbf{b}\|_2 + \lambda \cdot \|\mathbf{b}\|_1 \right\}, \quad (23)$$

where the penalty parameter $\lambda$ is defined as in (22) (but with $[\mathbf{X}^{(1)}_{\widehat{S}_0} \ \widetilde{\mathbf{X}}^{(1)}_{\widehat{S}_0}]$ in place of $\mathbf{X}$). For data recycling, we optimize the same problem but with $[\mathbf{X}_{\widehat{S}_0} \ \widetilde{\mathbf{X}}_{\widehat{S}_0}]$ and $\mathbf{y}$ in place of $[\mathbf{X}^{(1)}_{\widehat{S}_0} \ \widetilde{\mathbf{X}}^{(1)}_{\widehat{S}_0}]$ and $\mathbf{y}^{(1)}$. In each case, we then define $W_j = |\hat{\beta}_j| - |\tilde{\beta}_j|$ for each $j \in \widehat{S}_0$, and then apply the knockoff filter.

We run each method with $\kappa \in \{0.1, 0.2, \ldots, 1\}$ and find the highest power at $\kappa = 0.5$ for data splitting and $\kappa = 0.7$ for data recycling; only these values of $\kappa$ are displayed in our results.

## 5.3 Results

The knockoff filter and the BH procedure both work within the reduced model defined by the subset of features $\widehat{S}_0$ chosen at the initial screening step; the inference guarantees of these methods are with reference to this large submodel. On the other hand, the p-values computed by Tian et al. [30]'s method are calculated with reference to a relatively small submodel, given by features selected by the square-root Lasso. Therefore, in comparing these three types of methods against each other, it does not make sense to compare their FDRs over the partial regression coefficients, since these partial regression coefficients are defined within vastly different submodels. Instead, for each correlation setting and for each method tested, we compute the following measures of performance relative to the full model (averaged over 100 trials):

- False discovery rate and directional false discovery rate relative to the true full model: the average proportion of the selected features that were selected incorrectly (for FDR), or selected with an incorrect sign (for FDR$_{\text{dir}}$); and

- Power relative to the true full model: the average proportion of the true support, the $k_0 + k_1$ many features in $\text{support}(\boldsymbol{\beta})$, which is selected by the method. Since this true support is split into strong and weak signals, and the weak signals are barely detectable, we also report the "restricted power", that is, the power restricted to only the size-$k_0$ set of strong signals.

For all methods, the full-model FDR and directional FDR, along with the power and the restricted power, are displayed in Table ?? for both the AR design and the GWAS design settings. For both settings, we see that the methods all successfully control the FDR at the desired level, and indeed some are conservative, except that for large correlation values $\rho$ in the AR setting, the least squares + BH method does show overly high FDR (however, this can be explained by the fact that, for a highly correlated design, the FDR in the reduced model, i.e. for the partial regression coefficients $\beta^{\text{partial}}$, will be substantially different from the FDR in the full model). Across all settings and all methods, the directional FDR is slightly higher than the FDR, since the $k_1$ weak signals are extremely close to zero and, if selected, can easily be selected with the wrong sign. The (non-restricted) power is quite low for all methods, since the $k_1$ weak signals are extremely difficult to distinguish from zero. Turning instead to the restricted power, indicating each method's ability to detect strong signals, we see that the knockoff filter with data recycling shows somewhat higher power than the other methods, and in particular, we see a clear gain when using data recycling rather than data splitting, as expected.

---

[7]The software [29] sets $\lambda$ as $\kappa$ times the 95th percentile of $\frac{\|\mathbf{X}^\top \mathbf{g}\|_\infty}{\|\mathbf{g}\|_2}$ in lieu of what is diplayed in (22).

|                        |              | Least-sq. + BH | Selective inf. [30] | Knockoff/split | Knockoff/recycle |
|------------------------|--------------|----------------|---------------------|----------------|------------------|
| AR design ($\rho = 0$) | FDR          | 17.14 (0.7)    | 6.27 (0.5)          | 7.52 (0.5)     | 12.65 (0.8)      |
|                        | Dir. FDR     | 17.48 (0.7)    | 6.36 (0.5)          | 7.63 (0.5)     | 12.84 (0.8)      |
|                        | Power        | 9.47 (0.2)     | 9.67 (0.3)          | 8.85 (0.2)     | 11.08 (0.2)      |
|                        | Restr. power | 43.42 (1.1)    | 56.34 (2.0)         | 51.58 (1.0)    | 62.60 (1.0)      |
| AR design ($\rho = 0.25$) | FDR        | 16.93 (0.8)    | 6.46 (0.5)          | 8.72 (0.6)     | 12.11 (0.7)      |
|                        | Dir. FDR     | 17.31 (0.8)    | 6.51 (0.5)          | 8.84 (0.6)     | 12.27 (0.7)      |
|                        | Power        | 9.17 (0.2)     | 8.97 (0.4)          | 8.60 (0.2)     | 10.60 (0.2)      |
|                        | Restr. power | 41.66 (0.9)    | 51.98 (2.1)         | 49.96 (1.3)    | 60.56 (0.9)      |
| AR design ($\rho = 0.5$) | FDR         | 19.42 (0.9)    | 7.71 (0.6)          | 9.62 (0.8)     | 13.51 (1.0)      |
|                        | Dir. FDR     | 19.66 (0.9)    | 7.73 (0.6)          | 9.78 (0.8)     | 13.74 (1.0)      |
|                        | Power        | 7.81 (0.2)     | 8.75 (0.3)          | 6.65 (0.3)     | 8.96 (0.3)       |
|                        | Restr. power | 34.60 (1.0)    | 50.54 (1.9)         | 38.22 (1.7)    | 50.36 (1.4)      |
| AR design ($\rho = 0.75$) | FDR        | 29.99 (1.2)    | 13.72 (1.6)         | 14.15 (1.4)    | 18.25 (1.4)      |
|                        | Dir. FDR     | 30.43 (1.2)    | 14.02 (1.6)         | 14.28 (1.4)    | 18.40 (1.4)      |
|                        | Power        | 5.21 (0.2)     | 3.63 (0.3)          | 3.43 (0.3)     | 4.99 (0.3)       |
|                        | Restr. power | 20.10 (0.8)    | 21.00 (1.7)         | 19.38 (1.7)    | 27.90 (1.7)      |
| GWAS design            | FDR          | 17.97 (1.15)   | 8.62 (0.81)         | 12.28 (0.88)   | 15.05 (1.15)     |
|                        | Dir. FDR     | 18.36 (1.18)   | 9.04 (0.80)         | 12.56 (0.90)   | 15.41 (1.13)     |
|                        | Power        | 1.25 (0.05)    | 0.90 (0.05)         | 1.11 (0.04)    | 1.61 (0.05)      |
|                        | Restr. power | 22.08 (0.99)   | 24.24 (1.19)        | 26.76 (1.03)   | 37.20 (0.95)     |

**Table 1:** Simulation results. (Target FDR level is 20%.) Estimated standard errors are in parentheses.

## 6  Real Data Experiments

We next implement our high-dimensional knockoff method on a genome wide association study (GWAS) data set. The data comes from the Northern Finland Birth Cohort 1966 (NFBC1966) [26, 17], made available through the dbGaP database (accession number phs000276.v2.p1). Here, 5402 SNP arrays from subjects born in Northern Finland in 1966 are recorded, as well as a number of phenotype variables measured when the subjects were 31 years old.

### 6.1  Methods

**Data pre-processing.** Before applying knockoffs, the data needs to be prepared and our early pre-processing follows Sabatti et al. [26] and Janson et al. [16]. Genotype features from the original data set were removed if they met any of the following conditions:

- Not a SNP (some features were, e.g., copy number variations)
- Greater than 5% of values were missing
- All nonmissing values belonged to the same nucleotide
- SNP location could not be aligned to the genome
- A $\chi^2$ test rejected Hardy-Weinberg equilibrium at the 0.01% level
- On chromosome 23 (sex chromosome)

The remaining missing values were assumed to take the population frequency of the major allele. In the end, we have a total of $p = 328,934$ SNP features.

For each phenotype, we performed further processing on the response variables. Triglycerides were log-transformed. C-reactive protein (CRP) was also log-transformed after adding 0.002 mg/l (half the detection limit) to any values that were recorded as zero. Subjects were excluded from the triglycerides, HDL cholesterol, and LDL cholesterol if they were on diabetic medication or had not fasted before blood collection (or if either value was missing), or if they were pregnant, or if their respective phenotype measurement was more than three standard deviations from the mean, after correcting for sex, oral contraceptive use, and pregnancy. For the four phenotypes of interest to us, namely, CRP, HDL, LDL and triglycerides, the sample sizes are respectively 5290, 4700, 4682 and 4644. Due to space limitation, we report our findings on HDL and LDL only.

Finally, in order to correct for population stratification, we regressed the response and the features (SNPs) on the top five principal components of the design matrix in exactly the same fashion as suggested in Price et al. [25].

**Data processing with knockoffs.** Now that we have a response/design pair $(\mathbf{y}, \mathbf{X})$, we discuss the implementation of our full GWAS processing pipeline.

1. **Data splitting with random rotations.** In Section 4.1, we discuss splitting the data $(\mathbf{X}, \mathbf{y})$ into two parts, $(\mathbf{X}^{(0)}, \mathbf{y}^{(0)})$ and $(\mathbf{X}^{(1)}, \mathbf{y}^{(1)})$, with the first part used for screening and the second for inference; in that section, our discussion centers on splitting by partitioning the $n$ observations into two groups. In practice, for settings such as GWAS where the design matrix $\mathbf{X}$ is naturally somewhat sparse, this may be problematic as splitting the set of observations can dramatically increase conditioning problems caused by sparsity. Here we take a different approach to splitting. Under the Gaussian linear model, since the i.i.d. Gaussian noise term $\boldsymbol{\epsilon} \in \mathbb{R}^n$ has a distribution that is unchanged by rotation, we can rotate the data with a randomly chosen orthonormal matrix $\mathbf{U} \in \mathbb{R}^{n \times n}$, to obtain $(\mathbf{X}', \mathbf{y}') = (\mathbf{U}\mathbf{X}, \mathbf{U}\mathbf{y})$. The original linear model is preserved; as long as the rotation was chosen independently of the data, our model is now
$$\mathbf{y}' = \mathbf{X}'\boldsymbol{\beta} + \boldsymbol{\epsilon}',$$
where $\boldsymbol{\epsilon}' = \mathbf{U}\boldsymbol{\epsilon} \sim \mathcal{N}(0, \sigma^2 \mathbf{I}_n)$. (Note that $\boldsymbol{\beta}$ is unchanged from the original model.) After rotation, the new design matrix $\mathbf{X}'$ is now dense, and we are able to split the rotated data into a set of $n_0$ data points (for screening) and $n_1 = n - n_0$ data points (for inference) without matrix conditioning issues. From this point on, the partitioned data sets $(\mathbf{X}^{(0)}, \mathbf{y}^{(0)})$ and $(\mathbf{X}^{(1)}, \mathbf{y}^{(1)})$ are assumed to be taken from the rotated data, without further mention. We take $n_0 = 1900$.

2. **Feature pre-screening with correlations.** The original number of features (SNPs) is 328,934, which is a bit large to easily run the Lasso for feature screening. As suggested in [37] (where the screen-and-clean methodology is applied to GWAS), we begin with a pre-screening step that looks only at marginal correlations: using the first part of the data $(\mathbf{X}^{(0)}, \mathbf{y}^{(0)})$, we select the set $\widehat{S}_{\text{pre}} \subset [p]$ of 26,300 features with the largest magnitude correlations $\left| \mathbf{X}_j^{(0)\top} \mathbf{y}^{(0)} \right|$.

3. **Feature screening with Lasso.** We then run the Lasso on $(\mathbf{X}_{\widehat{S}_{\text{pre}}}^{(0)}, \mathbf{y}^{(0)})$ using the 26,300 pre-screened features as covariates, and define the screened set of features $\widehat{S}_0$ to be the first $n_1/4$ features, which enter the Lasso path on this reduced data set.

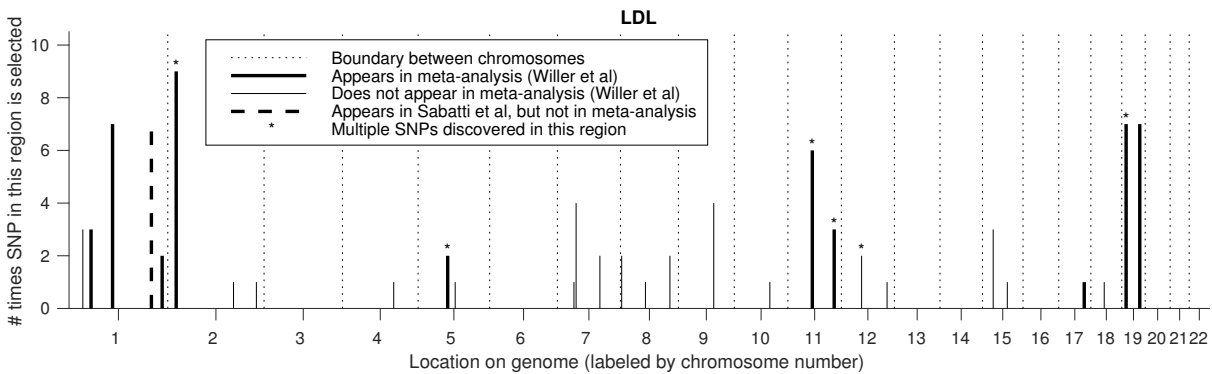

**Figure 1:** Results of GWAS experiment for the LDL phenotype. The heights of the lines show the number of trials (out of 10 trials) for which at least one SNP in this region was selected; the line type shows whether this same region was identified in [36] and/or [26]. Regions where more than one SNP was selected (across the 10 trials) are marked with a '*'. (See Section 6.2 for details.)

4. **Knockoff filter.** Next we bring back the held out part of the data, $(\mathbf{X}^{(1)}, \mathbf{y}^{(1)})$. We apply the knockoff filter with recycling as defined in Section 4.1, with statistics determined by the sign-restricted Lasso introduced in Section 4.1.2. Specifically,, we fit the sign-restricted Lasso path

$$\min_{\mathbf{b} \in \mathbb{R}^{2|\widehat{S}_0|}} \quad \tfrac{1}{2}\|\mathbf{y} - [\mathbf{X}_{\widehat{S}_0} \; \widetilde{\mathbf{X}}_{\widehat{S}_0}]\mathbf{b}\|_2^2 + \lambda\|\mathbf{b}\|_1$$
$$\text{s.t.} \quad b_j \cdot \text{sign}_j^{(0)} \geq 0$$
$$\quad b_{j+|\widehat{S}_0|} \cdot \text{sign}_j^{(0)} \geq 0$$

and set $W_j$ as in (9). (That is, we record the largest $\lambda$ such that $\mathbf{X}_j$ or $\widetilde{\mathbf{X}}_j$ enters the Lasso path, and set $W_j = +\lambda$ if $\mathbf{X}_j$ enters first, and $W_j = -\lambda$ otherwise.) Throughout, we use target FDR level $q = 0.2$.

5. **Repeat.** We then repeat all these steps, with a new random rotation matrix $\mathbf{U} \in \mathbb{R}^{n \times n}$, for a total of 10 repetitions. In particular, we are interested in examining whether any selected SNPs appear consistently across these 10 repetitions of our method.

## 6.2   Results

Our results for the LDL phenotype are displayed in Figure 1. (The full details of the results, labeled according to each individual SNP discovered in the analysis, are reported in Appendix C.) In total, over the 10 trials (the random splits of the data), 44 different SNPs are selected. However, as nearby SNPs are extremely correlated, we consider SNPs whose positions are within $10^6$ base pairs of each other, to be in the same "region"; at this level, in total over the 10 trials, there are 29 distinct regions discovered. In Figure 1 we show, for each region, the selection frequency of any SNP in that region, i.e. the number of trials out of 10 for which at least one SNP in this region was selected (we also display the selection frequency for the individual SNPs). We also show any SNPs in the same region (defined again as a distance of $\leq 10^6$ base pairs) which were identified as associated with LDL in the meta-analysis of Willer et al. [36]; this meta-analysis works with an extremely large sample size, and therefore can be viewed as providing a form of "ground truth". Comparing against the findings of this meta-analysis, we can try to estimate the FDR of our method as follows: for each trial, we count each selected SNP as a discovery, and label it as a false discovery if there is no SNP in the meta-analysis from the same region. We then average this false discovery proportion over the 10 trials, and find an estimated FDR of 32.61%. We also compare against the findings of Sabatti et al. [26], which analyzes the same data set that we use; this comparison therefore should not be viewed as independent validation, but rather as checking that our analysis agrees with existing work on the same data set.

For the HDL phenotype, our results are shown in Figure 2 (and reported in full detail in Appendix C). There are 13 SNPs discovered at least once in the 10 trials; once grouped into regions, there are 10 distinct regions. Again estimating the FDR by comparing against Willer et al. [36]'s meta-analysis, we estimate an FDR of 4.29%. This low number of discoveries, and perhaps conservative FDR, can probably be attributed to fact that this data set has a relatively small sample size, and fairly weak signal; we would not expect a large number of discoveries in this setting.

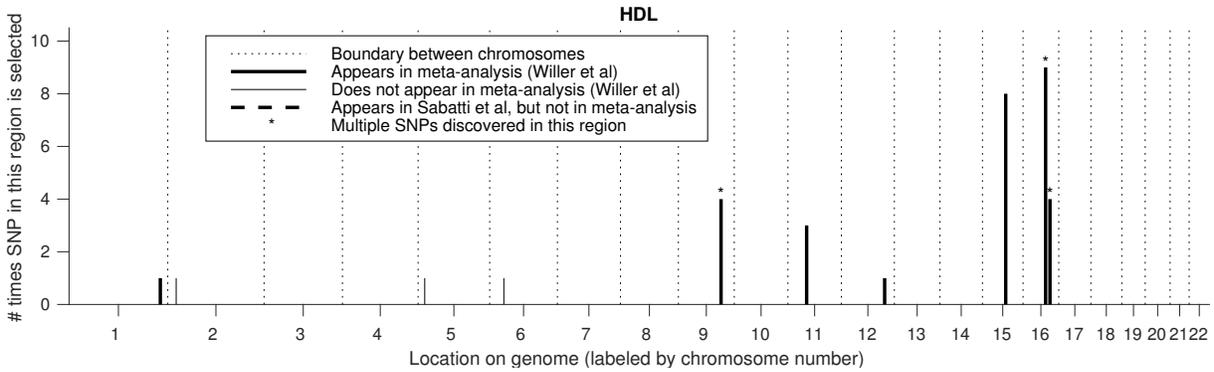

**Figure 2:** Results of GWAS experiment for the HDL phenotype (same interpretation as in Figure 1).

# 7  Summary

We have discussed a 'screen + knockoff' approach to controlled variable selection, which should appear very natural to practitioners; in addition, we develop the "recycling" technique to reuse data rather than discarding the data used for the screening step, thus improving power. A natural question is thus what sort of inference properties would such a procedure offer? By focusing on Type I + Type S errors, we have shown that screen + knockoff is guaranteed to have a form of reproducibility in that the directional FDR is rigorously under control. This result makes no assumption on the design matrix nor on the values of the regression coefficients, and holds in finite samples; our contribution is, therefore, very different in spirit from most of the results published in the literature on high-dimensional inference (the work of Taylor and colleagues referenced earlier being a notable exception), and employs mathematical arguments centered in martingale theory which are naturally also very different from the techniques used in the related literature.

### Acknowledgements

E. C. was partially supported by the NSF via grant CCF-0963835 and by the Math + X Award from the Simons Foundation. R. F. B. was partially supported by the NSF via grant DMS-1654076 and by an Alfred P. Sloan fellowship. E. C. would like to thank Lucas Janson, Chiara Sabatti and Jonathan Taylor for many inspiring discussions about topics ranging from high-dimensional inference to GWAS, and useful comments about an early version of the manuscript. We are indebted to Chiara Sabatti and Christine Peterson for all their help with the NFBC data set and its pre-processing, and to Pragya Sur for her editorial comments and assistance. E. C. would like to thank Mert Pilanci and Zhimei Ren for their help in running simulations on the Sherlock cluster.

# A Proofs

## A.1 A general form of directional FDR control

Our results, namely, Theorems 1–3, are all special cases of the general statement below. Throughout, $\widehat{\text{sign}}_j$ is defined as in (15), i.e. $\widehat{\text{sign}}_j = \text{sign}\left((\mathbf{X}_j - \widetilde{\mathbf{X}}_j)^\top \mathbf{y}\right)$. Also, we shall say that a positive semidefinite matrix $\mathbf{M} \in \mathbb{R}^{2p \times 2p}$ satisfies the *pairwise exchangeability condition* if $M_{j,k} = M_{j,k+p} = M_{j+p,k} = M_{j+p,k+p}$ for all $j \neq k \in \{1, \ldots, p\}$ and $M_{j,j} = M_{j+p,j+p}$ for all $j \in \{1, \ldots, p\}$.

**Theorem 4.** *Suppose $\mathbf{y} \sim \mathcal{N}(\boldsymbol{\mu}, \boldsymbol{\Theta})$ for some mean vector $\boldsymbol{\mu} \in \mathbb{R}^n$ and some covariance $\boldsymbol{\Theta} \in \mathbb{R}^{n \times n}$. Let $\mathbf{X}, \widetilde{\mathbf{X}} \in \mathbb{R}^{n \times p}$ be any fixed matrices such that $\begin{bmatrix} \mathbf{X} & \widetilde{\mathbf{X}} \end{bmatrix}^\top \cdot \boldsymbol{\Theta} \cdot \begin{bmatrix} \mathbf{X} & \widetilde{\mathbf{X}} \end{bmatrix}$ obeys the pairwise exchangeability condition. Let $\mathbf{W} = (W_1, \ldots, W_p)$ be any statistic satisfying the sufficiency and antisymmetry properties (7) and (8). Finally, let $V_j = \mathbb{1}\left\{\widehat{\text{sign}}_j \neq \text{sign}((\mathbf{X}_j - \widetilde{\mathbf{X}}_j)^\top \boldsymbol{\mu})\right\}$ be the indicator of a sign error on feature $j$. Then the knockoff+ method controls a directional FDR defined as*

$$\text{FDR}_{\text{dir}} = \mathbb{E}\left[\frac{\left|\left\{j \in \widehat{S} \text{ and } V_j = 1\right\}\right|}{1 \vee |\widehat{S}|}\right] \leq q\,.$$

*If the knockoff method is used instead, then a modified directional FDR is controlled:*

$$\text{mFDR}_{\text{dir}} = \mathbb{E}\left[\frac{\left|\left\{j \in \widehat{S} \text{ and } V_j = 1\right\}\right|}{|\widehat{S}| + q^{-1}}\right] \leq q\,.$$

We prove this theorem in Appendix B. The rest of this section shows how all of our results follow as special cases.

**Proof of Theorem 1** This is straightforward since the result is obtained by taking $\boldsymbol{\mu} = \mathbf{X}\boldsymbol{\beta}$ and $\boldsymbol{\Theta} = \sigma^2 \mathbf{I}_n$ in the distribution of $\mathbf{y}$, and noting that $\text{sign}((\mathbf{X}_j - \widetilde{\mathbf{X}}_j)^\top \boldsymbol{\mu}) = \text{sign}(\beta_j)$ for all $j$ such that $\mathbf{X}_j \neq \widetilde{\mathbf{X}}_j$. (We ignore the trivial case where $s_j = 0$ implying $\mathbf{X}_j = \widetilde{\mathbf{X}}_j$, for in this case we would get $W_j = 0$ and thus feature $j$ could never be selected.)

The more general form of Theorem 4 also allows us to consider other settings, such as those involving variable screening in high dimensions. For instance, our next proof concerns the testing of regression coefficients in the full model under the sure screening property.

**Proof of Theorem 2** We prove directional FDR control by conditioning on $\mathbf{y}^{(0)}$. For any $\mathbf{y}^{(0)}$ such that the event $\mathcal{E}$ holds, we will show that for the knockoff procedure,

$$\mathbb{E}\left[\text{mFDR}_{\text{dir}} \,\Big|\, \mathbf{y}^{(0)}\right] \cdot \mathbb{1}_{\mathcal{E}} \leq q\,, \tag{24}$$

and for the knockoff+,

$$\mathbb{E}\left[\text{FDR}_{\text{dir}} \,\Big|\, \mathbf{y}^{(0)}\right] \cdot \mathbb{1}_{\mathcal{E}} \leq q\,. \tag{25}$$

From this point on we treat $\mathbf{y}^{(0)}$ as fixed. The conditional distribution of the response $\mathbf{y}$ is given by

$$(\mathbf{y} \mid \mathbf{y}^{(0)}) \sim \mathcal{N}\left(\begin{bmatrix} \mathbf{y}^{(0)} \\ \mathbf{X}^{(1)}\boldsymbol{\beta} \end{bmatrix}, \begin{bmatrix} 0 & 0 \\ 0 & \sigma^2 \mathbf{I}_{n_1} \end{bmatrix}\right) . \tag{26}$$

If $\mathbf{y}^{(0)}$ is such that the event $\mathcal{E}$ occurs, then we can simplify this to

$$(\mathbf{y} \mid \mathbf{y}^{(0)}) \sim \mathcal{N}(\boldsymbol{\mu}, \boldsymbol{\Theta}), \text{ where } \boldsymbol{\mu} = \begin{bmatrix} \mathbf{y}^{(0)} \\ \mathbf{X}^{(1)}_{\widehat{S}_0}\boldsymbol{\beta}_{\widehat{S}_0} \end{bmatrix} \text{ and } \boldsymbol{\Theta} = \begin{bmatrix} 0 & 0 \\ 0 & \sigma^2 \mathbf{I}_{n_1} \end{bmatrix} , \tag{27}$$

since $\beta_j = 0$ for all $j \notin \widehat{S}_0$ when $\mathcal{E}$ occurs. Define $\boldsymbol{\Sigma}^{(1)}_{\widehat{S}_0} = \mathbf{X}^{(1)\top}_{\widehat{S}_0}\mathbf{X}^{(1)}_{\widehat{S}_0}$. Recall that $\widetilde{\mathbf{X}}^{(0)}_{\widehat{S}_0} = \mathbf{X}^{(0)}_{\widehat{S}_0}$ by (17), and that $\widetilde{\mathbf{X}}^{(1)}_{\widehat{S}_0}$ satisfies

$$[\mathbf{X}^{(1)}_{\widehat{S}_0}\ \widetilde{\mathbf{X}}^{(1)}_{\widehat{S}_0}]^\top [\mathbf{X}^{(1)}_{\widehat{S}_0}\ \widetilde{\mathbf{X}}^{(1)}_{\widehat{S}_0}] = \begin{bmatrix} \boldsymbol{\Sigma}^{(1)}_{\widehat{S}_0} & \boldsymbol{\Sigma}^{(1)}_{\widehat{S}_0} - \text{diag}\{\mathbf{s}\} \\ \boldsymbol{\Sigma}^{(1)}_{\widehat{S}_0} - \text{diag}\{\mathbf{s}\} & \boldsymbol{\Sigma}^{(1)}_{\widehat{S}_0} \end{bmatrix} \tag{28}$$

for some vector $\mathbf{s} \geq 0$, by the knockoff construction. We then see that

$$[\mathbf{X}_{\widehat{S}_0}\ \widetilde{\mathbf{X}}_{\widehat{S}_0}]^\top \boldsymbol{\Theta} [\mathbf{X}_{\widehat{S}_0}\ \widetilde{\mathbf{X}}_{\widehat{S}_0}] = \sigma^2 [\mathbf{X}^{(1)}_{\widehat{S}_0}\ \widetilde{\mathbf{X}}^{(1)}_{\widehat{S}_0}]^\top [\mathbf{X}^{(1)}_{\widehat{S}_0}\ \widetilde{\mathbf{X}}^{(1)}_{\widehat{S}_0}] ,$$

is a pairwise exchangeable matrix. At this point, we have satisfied the conditions of Theorem 4. Now what is a sign error in Theorem 4? We calculate that for any $j \in \widehat{S}_0$,

$$(\mathbf{X}_j - \widetilde{\mathbf{X}}_j)^\top \boldsymbol{\mu} = (\mathbf{X}^{(1)}_j - \widetilde{\mathbf{X}}^{(1)}_j)^\top \mathbf{X}^{(1)}_{\widehat{S}_0}\boldsymbol{\beta}_{\widehat{S}_0} = \sum_{k \in \widehat{S}_0}(\mathbf{X}^{(1)}_j - \widetilde{\mathbf{X}}^{(1)}_j)^\top \mathbf{X}^{(1)}_k \cdot \beta_k = s_j \cdot \beta_j ,$$

where the first equality holds on the event $\mathcal{E}$ (i.e. $\beta_k = 0$ for $k \notin \widehat{S}_0$), and the third equality uses the Gram matrix condition (28). Hence, since $s_j > 0$ whenever $\mathbf{X}_j \neq \widetilde{\mathbf{X}}_j$, an error $\widehat{\text{sign}}_j \neq \text{sign}((\mathbf{X}_j - \widetilde{\mathbf{X}}_j)^\top \boldsymbol{\mu})$ is the same as $\widehat{\text{sign}}_j \neq \text{sign}(\beta_j)$, which completes the proof.

We now move on to our result concerning regression coefficients from the reduced model.

**Proof of Theorem 3** This last proof is more subtle and for this, we treat $\mathbf{X}$ as random in addition to $\mathbf{y}$, but will condition on $\mathbf{X}^{(0)}$ and on the screened features $\mathbf{X}^{(1)}_{\widehat{S}_0}$ and knockoffs $\widetilde{\mathbf{X}}^{(1)}_{\widehat{S}_0}$ as well as on $\mathbf{y}^{(0)}$. By the assumption that the rows of $\mathbf{X}$ are i.i.d. draws from $\mathcal{N}(0, \boldsymbol{\Psi})$, we can write

$$\mathbf{X}^{(1)}_{\widehat{S}_0^c} = \mathbf{X}^{(1)}_{\widehat{S}_0} \cdot \boldsymbol{\Gamma} + \mathbf{G} \cdot \boldsymbol{\Lambda}^{1/2} ,$$

where $\boldsymbol{\Gamma}$ and $\boldsymbol{\Lambda}$ are the fixed unknown matrices given by

$$\boldsymbol{\Gamma} = (\boldsymbol{\Psi}_{\widehat{S}_0, \widehat{S}_0})^{-1} \cdot \boldsymbol{\Psi}_{\widehat{S}_0, \widehat{S}_0^c}$$

and the Schur complement

$$\boldsymbol{\Lambda} = \boldsymbol{\Psi}_{\widehat{S}_0^c, \widehat{S}_0^c} - \boldsymbol{\Psi}_{\widehat{S}_0^c, \widehat{S}_0} \cdot (\boldsymbol{\Psi}_{\widehat{S}_0, \widehat{S}_0})^{-1} \cdot \boldsymbol{\Psi}_{\widehat{S}_0, \widehat{S}_0^c} ,$$

and where $\mathbf{G} \in \mathbb{R}^{n_1 \times (p - |\widehat{S}_0|)}$ has i.i.d. standard normal entries drawn independently from $(\mathbf{X}^{(0)}, \mathbf{y}^{(0)})$ and from $(\mathbf{X}^{(1)}_{\widehat{S}_0}, \widetilde{\mathbf{X}}^{(1)}_{\widehat{S}_0}, \boldsymbol{\epsilon}^{(1)})$. Then

$$\mathbf{y}^{(1)} = \mathbf{X}^{(1)}\boldsymbol{\beta} + \boldsymbol{\epsilon}^{(1)} = \mathbf{X}^{(1)}_{\widehat{S}_0} \cdot \left(\boldsymbol{\beta}_{\widehat{S}_0} + \boldsymbol{\Gamma}\boldsymbol{\beta}_{\widehat{S}_0^c}\right) + \mathbf{G} \cdot \boldsymbol{\Lambda}^{1/2}\boldsymbol{\beta}_{\widehat{S}_0^c} + \boldsymbol{\epsilon}^{(1)} ,$$

and note that

$$\boldsymbol{\beta}_{\widehat{S}_0} + \boldsymbol{\Gamma}\boldsymbol{\beta}_{\widehat{S}_0^c} = \boldsymbol{\beta}^{\text{partial}}$$

by definition of the expected partial regression coefficients. This is because from (20), we get that

$$\boldsymbol{\beta}^{\text{partial}} = \mathbf{X}_{\widehat{S}_0}^{(1)\dagger}\left(\mathbf{X}_{\widehat{S}_0}^{(1)}\boldsymbol{\beta}_{\widehat{S}_0} + \mathbb{E}\left[\mathbf{X}_{\widehat{S}_0^c}^{(1)}\boldsymbol{\beta}_{\widehat{S}_0^c} \,\Big|\, \mathbf{X}_{\widehat{S}_0}^{(1)}; \mathbf{X}^{(0)}, \mathbf{y}^{(0)}\right]\right) = \mathbf{X}_{\widehat{S}_0}^{(1)\dagger}\left(\mathbf{X}_{\widehat{S}_0}^{(1)}\boldsymbol{\beta}_{\widehat{S}_0} + \mathbf{X}_{\widehat{S}_0}^{(1)}\boldsymbol{\Gamma}\boldsymbol{\beta}_{\widehat{S}_0^c}\right)$$
$$= \boldsymbol{\beta}_{\widehat{S}_0} + \boldsymbol{\Gamma}\boldsymbol{\beta}_{\widehat{S}_0^c},$$

where $\mathbf{X}_{\widehat{S}_0}^{(1)\dagger}$ is a short hand for $(\mathbf{X}_{\widehat{S}_0}^{(1)\top}\mathbf{X}_{\widehat{S}_0}^{(1)})^{-1}\mathbf{X}_{\widehat{S}_0}^{(1)\top}$. Therefore,

$$(\mathbf{y} \mid \mathbf{X}^{(0)}, \mathbf{y}^{(0)}, \mathbf{X}_{\widehat{S}_0}^{(1)}, \widetilde{\mathbf{X}}_{\widehat{S}_0}^{(1)}) \sim \mathcal{N}\left(\begin{bmatrix} \mathbf{y}^{(0)} \\ \mathbf{X}_{\widehat{S}_0}^{(1)}\boldsymbol{\beta}^{\text{partial}} \end{bmatrix}, \begin{bmatrix} 0 & 0 \\ 0 & \left(\boldsymbol{\beta}_{\widehat{S}_0^c}^{\top}\boldsymbol{\Lambda}\boldsymbol{\beta}_{\widehat{S}_0^c} + \sigma^2\right)\mathbf{I}_{n_1} \end{bmatrix}\right).$$

We are now in the same setting as in the proof of Theorem 2 (see (26)), except with $\boldsymbol{\beta}^{\text{partial}}$ in place of $\boldsymbol{\beta}$ and with a new variance level that has increased due to the randomness in the missed signals. The remainder of the proof thus proceeds identically.

# B  Additional proofs

In this section, we first prove Lemma 1 (presented in Section 3 of the main paper) and some additional supporting lemmas, and then present the proof of our general error control result, Theorem 4 (presented in Appendix A.1 of the main paper).

## B.1  Key lemmas

The proof of Theorem 4 relies on two key lemmas: Lemma 1 earlier which treats non-i.i.d. Bernoulli sequences, and the following result regarding the statistics $\mathbf{W}$ arising in the knockoff procedure.

**Lemma 2.** *With the assumptions from Theorem 4, let $\mathcal{V}$ be the $\sigma$-algebra generated by the random variables $\{(\mathbf{X} + \widetilde{\mathbf{X}})^{\top}\mathbf{y}, |(\mathbf{X} - \widetilde{\mathbf{X}})^{\top}\mathbf{y}|\}$, and define the vector $\mathbf{S} = (S_1, \ldots, S_p)$ of signs as*

$$S_j = \text{sign}\left((\mathbf{X}_j - \widetilde{\mathbf{X}}_j)^{\top}\mathbf{y}\right) \cdot \text{sign}(W_j).$$

*Then the following holds:*

- *$|\mathbf{W}|$ and $\mathbf{S}$ belong to $\mathcal{V}$.*

- *The signs of $W_j$ are mutually independent after conditioning on $\mathcal{V}$.*

- *For each $j = 1, \ldots, p$, if $S_j \neq \text{sign}\left((\mathbf{X}_j - \widetilde{\mathbf{X}}_j)^{\top}\boldsymbol{\mu}\right)$, then*

$$\mathbb{P}\left\{\text{sign}(W_j) = -1 \mid \mathcal{V}\right\} \geq 1/2.$$

Before presenting the formal proof, we give an intuitive explanation for why the probabilities for $\text{sign}(W_j)$ work in our favor; that is, why we are able to obtain $\mathbb{P}\left\{\text{sign}(W_j) = -1 \mid \mathcal{V}\right\} \geq 1/2$, which ensures that we are more likely to over- rather than under-estimate the number of false positives. This is similar to the discussion for the low-dimensional setting as in (15). In this case, for the high-dimensional setting, we see that if $S_j \neq \text{sign}\left((\mathbf{X}_j - \widetilde{\mathbf{X}}_j)^{\top}\boldsymbol{\mu}\right)$, then:

- If $(\mathbf{X}_j - \widetilde{\mathbf{X}}_j)^{\top}\boldsymbol{\mu} \geq 0$, then $\text{sign}(W_j) = -1$ whenever $(\mathbf{X}_j - \widetilde{\mathbf{X}}_j)^{\top}\mathbf{y} > 0$;

- If $(\mathbf{X}_j - \widetilde{\mathbf{X}}_j)^{\top}\boldsymbol{\mu} \leq 0$, then $\text{sign}(W_j) = -1$ whenever $(\mathbf{X}_j - \widetilde{\mathbf{X}}_j)^{\top}\mathbf{y} < 0$.

Of course, $(\mathbf{X}_j - \widetilde{\mathbf{X}}_j)^{\top}\mathbf{y}$ is a normal random variable whose mean is given by $(\mathbf{X}_j - \widetilde{\mathbf{X}}_j)^{\top}\boldsymbol{\mu}$. If this mean is nonnegative, then the random variable $(\mathbf{X}_j - \widetilde{\mathbf{X}}_j)^{\top}\mathbf{y}$ is more likely to be positive than negative; if this mean is nonpositive, then the opposite is true. Either way, we see that $\text{sign}(W_j) = -1$ has probability at least $1/2$.

*Proof.* First, by the sufficiency property, $\mathbf{W}$ is a function of $[\mathbf{X}\ \widetilde{\mathbf{X}}]^\top[\mathbf{X}\ \widetilde{\mathbf{X}}]$ and $[\mathbf{X}\ \widetilde{\mathbf{X}}]^\top\mathbf{y}$ or, equivalently, is of the form

$$\mathbf{W} = \mathbf{w}\left([\mathbf{X}\ \widetilde{\mathbf{X}}]^\top[\mathbf{X}\ \widetilde{\mathbf{X}}], (\mathbf{X}+\widetilde{\mathbf{X}})^\top\mathbf{y}, |(\mathbf{X}-\widetilde{\mathbf{X}})^\top\mathbf{y}|, \text{sign}((\mathbf{X}-\widetilde{\mathbf{X}})^\top\mathbf{y})\right)$$

for some function $\mathbf{w}$. Second, the antisymmetry property implies that flipping the signs of any subset of the last argument of the function $\mathbf{w}$ flips the signs of the corresponding outputs. (Of course, for any $j$ such that $(\mathbf{X}_j - \widetilde{\mathbf{X}}_j)^\top\mathbf{y} = 0$, the antisymmetry property implies that $W_j = 0$; we ignore these features in our proof from this point on as they can never be selected by our method.) This property makes clear that $\mathcal{V}$ determines $\mathbf{W}$ up to a sign change, so that $|\mathbf{W}| \in \mathcal{V}$. Furthermore, define a vector of signs $\mathbf{S}$ such that

$$|\mathbf{W}| = \mathbf{w}\left([\mathbf{X}\ \widetilde{\mathbf{X}}]^\top[\mathbf{X}\ \widetilde{\mathbf{X}}], (\mathbf{X}+\widetilde{\mathbf{X}})^\top\mathbf{y}, |(\mathbf{X}-\widetilde{\mathbf{X}})^\top\mathbf{y}|, \mathbf{S}\right).$$

Clearly, $\mathbf{S} \in \mathcal{V}$ and, by the antisymmetry property, we also see that

$$\mathbf{w}\left([\mathbf{X}\ \widetilde{\mathbf{X}}]^\top[\mathbf{X}\ \widetilde{\mathbf{X}}], (\mathbf{X}+\widetilde{\mathbf{X}})^\top\mathbf{y}, |(\mathbf{X}-\widetilde{\mathbf{X}})^\top\mathbf{y}|, \mathbf{S}\cdot\text{sign}(\mathbf{W})\right) = \mathbf{W}.$$

This equality means that $\mathbf{S}\cdot\text{sign}(\mathbf{W})$ must be equal to $\text{sign}((\mathbf{X}-\widetilde{\mathbf{X}})^\top\mathbf{y})$ or, expressed differently, $\mathbf{S} = \text{sign}(\mathbf{W})\cdot\text{sign}((\mathbf{X}-\widetilde{\mathbf{X}})^\top\mathbf{y})$ thereby coinciding with the definition given in the statement of the lemma. This establishes the first claim.

Next, an elementary calculation shows that

$$\begin{bmatrix}(\mathbf{X}+\widetilde{\mathbf{X}})^\top\mathbf{y}\\(\mathbf{X}-\widetilde{\mathbf{X}})^\top\mathbf{y}\end{bmatrix} \sim \mathcal{N}\left(\begin{bmatrix}(\mathbf{X}+\widetilde{\mathbf{X}})^\top\boldsymbol{\mu}\\(\mathbf{X}-\widetilde{\mathbf{X}})^\top\boldsymbol{\mu}\end{bmatrix}, \begin{bmatrix}(\mathbf{X}+\widetilde{\mathbf{X}})^\top\boldsymbol{\Theta}(\mathbf{X}+\widetilde{\mathbf{X}}) & 0 \\ 0 & \text{diag}\{\mathbf{d}\}\end{bmatrix}\right), \tag{29}$$

where $\mathbf{d} = (d_1,\ldots,d_p) \in \mathbb{R}^p$ is some nonnegative vector. The form of the covariance term arises from the fact that $(\mathbf{X}-\widetilde{\mathbf{X}})^\top\boldsymbol{\Theta}(\mathbf{X}+\widetilde{\mathbf{X}}) = 0$ and $(\mathbf{X}-\widetilde{\mathbf{X}})^\top\boldsymbol{\Theta}(\mathbf{X}-\widetilde{\mathbf{X}})$ is diagonal, according to the pairwise exchangeability assumption on $[\mathbf{X}\ \widetilde{\mathbf{X}}]^\top\boldsymbol{\Theta}[\mathbf{X}\ \widetilde{\mathbf{X}}]$. Therefore, the terms $(\mathbf{X}_j - \widetilde{\mathbf{X}}_j)^\top\mathbf{y}$ are mutually independent, and are independent from $(\mathbf{X}+\widetilde{\mathbf{X}})^\top\mathbf{y}$. Also, since $\mathbf{S} \in \mathcal{V}$ the terms $\text{sign}(W_j) = S_j \cdot \text{sign}((\mathbf{X}_j - \widetilde{\mathbf{X}}_j)^\top\mathbf{y})$ are mutually independent after conditioning on $\mathcal{V}$, proving our second claim.

Finally, observe that if $Z \sim \mathcal{N}(\theta, \sigma^2)$, then it follows from a simple calculation that

$$\frac{\mathbb{P}\{\text{sign}(Z) = -1 \mid |Z|\}}{\mathbb{P}\{\text{sign}(Z) = +1 \mid |Z|\}} = \exp\left(-\frac{2\theta|z|}{\sigma^2}\right).$$

Now since $\mathbb{P}\{\text{sign}(W_j) = \pm 1 \mid \mathcal{V}\} = \mathbb{P}\{S_j \cdot \text{sign}((\mathbf{X}_j - \widetilde{\mathbf{X}}_j)^\top\mathbf{y}) = \pm 1 \mid \mathcal{V}\}$, applying the above formula gives

$$\frac{\mathbb{P}\{\text{sign}(W_j) = -1 \mid \mathcal{V}\}}{\mathbb{P}\{\text{sign}(W_j) = +1 \mid \mathcal{V}\}} = \exp\left(-\frac{2S_j \cdot (\mathbf{X}_j - \widetilde{\mathbf{X}}_j)^\top\boldsymbol{\mu} \cdot |(\mathbf{X}_j - \widetilde{\mathbf{X}}_j)^\top\mathbf{y}|}{d_j}\right).$$

Clearly, if $S_j \neq \text{sign}\left((\mathbf{X}_j - \widetilde{\mathbf{X}}_j)^\top\boldsymbol{\mu}\right)$, then the argument in the exponential in the right-hand side of this last inequality is positive and thus, $\mathbb{P}\{\text{sign}(W_j) = -1 \mid \mathcal{V}\} \geq 1/2$. $\square$

The original knockoff paper [1] worked in the simpler setting where the signs of $W_j$ for the nulls are i.i.d. unbiased. If $\mathcal{H}_0$ is the set of nulls and $T > 0$ the adaptive threshold of the knockoff filter, [1] established that the i.i.d. property for the signs gives

$$\mathbb{E}\left[\mathbb{E}\left[\frac{|\{j \in \mathcal{H}_0 : W_j \geq T\}|}{1 + |\{j \in \mathcal{H}_0 : W_j \leq -T\}|} \,\bigg|\, |\mathbf{W}|\right]\right] \leq 1.$$

Here, we use Lemma 1 to develop a similar bound when the signs are no longer i.i.d. unbiased but are, instead, independent with at most a 50% chance of being positive conditionally on $\mathcal{V}$.

**Corollary 1.** *Suppose $\mathcal{M} \subset \{1, \ldots, p\}$ is a random set belonging to $\mathcal{V}$ and chosen such that for each $j \in \mathcal{M}$, $|W_j| > 0$ and $\mathbb{P}\{\text{sign}(W_j) = -1 \mid \mathcal{V}\} \geq \rho$. Letting $T > 0$ be the adaptive threshold of either the knockoff or knockoff+ filter, then*

$$\mathbb{E}\left[\frac{|\{j \in \mathcal{M} : W_j \geq T\}|}{1 + |\{j \in \mathcal{M} : W_j \leq -T\}|} \, \bigg| \, \mathcal{V}\right] \leq \rho^{-1} - 1. \tag{30}$$

*Proof.* The proof is straightforward and only consists in rewriting the left-hand side of (30) in such a way that we recognize the formulation from Lemma 1. To begin with, we can treat the ordering of $|\mathbf{W}|$ as fixed since Lemma 2 gives $|\mathbf{W}| \in \mathcal{V}$. Now assume without loss of generality that $\mathcal{M} = \{1, \ldots, m\}$ and reorder the indices of the $W_j$'s in $\mathcal{M}$ so that $|W_{(1)}| \geq |W_{(2)}| \geq \cdots \geq |W_{(m)}|$. Set $B_j = \mathbb{1}_{W_{(j)} < 0}$. Next we will condition also on $\text{sign}(W_j)$ for all $j \notin \mathcal{M}$. Let $\widetilde{\mathcal{V}}$ be the larger $\sigma$-algebra generated by $\mathcal{V}$ and $\{\text{sign}(W_j) : j \notin \mathcal{M}\}$. By definition of $\mathcal{M}$, when we condition on $\mathcal{V}$, we have $B_j \stackrel{\perp}{\sim} \text{Bernoulli}(\rho_j)$ with $\rho_j \geq \rho$ for all $j \in \mathcal{M}$. By Lemma 2, the same is still true even after conditioning on $\{\text{sign}(W_j) : j \notin \mathcal{M}\}$ also, since the signs of $\mathbf{W}$ are mutually independent conditional on $\mathcal{V}$. To summarize, we have

$$(B_j)_{j \in \mathcal{M}} \mid \widetilde{\mathcal{V}} \stackrel{\perp}{\sim} \text{Bernoulli}(\rho_j),$$

with $\rho_j \geq \rho$ for all $j \in \mathcal{M}$.

Next, we can write

$$\frac{\#\{j \in \mathcal{M} : W_j \geq T\}}{1 + \#\{j \in \mathcal{M} : W_j \leq -T\}} = \frac{(1 - B_1) + \cdots + (1 - B_J)}{1 + B_1 + \cdots + B_J} = \frac{1 + J}{1 + B_1 + \cdots + B_J} - 1,$$

where $J$ is the index such that $|W_{(1)}| \geq \cdots \geq |W_{(J)}| \geq T > |W_{(J+1)}| \geq \cdots \geq |W_{(m)}|$. After conditioning on $\widetilde{\mathcal{V}}$, so that we can treat $|\mathbf{W}|$ and $\{\text{sign}(W_j) : j \notin \mathcal{M}\}$ as fixed, we observe that the index $J$ can be expressed as a stopping time, in reverse time, with respect to the filtration $\{\mathcal{F}_j\}$ given by

$$\mathcal{F}_j = \{B_1 + \cdots + B_j, B_{j+1}, \ldots, B_m\}.$$

The conclusion now follows from Lemma 1. □

## B.2 Proof of Theorem 4

First, consider control of mFDR$_{\text{dir}}$ under the knockoff. Since $j \in \widehat{S}$ if and only if $W_j \geq T$ where $T$ is our adaptive threshold, the modified directional FDR is given by

$$\text{mFDR}_{\text{dir}} = \mathbb{E}\left[\frac{\left|\left\{j : W_j \geq T, \widehat{\text{sign}}_j \neq \text{sign}((\mathbf{X}_j - \widetilde{\mathbf{X}}_j)^\top \boldsymbol{\mu})\right\}\right|}{|\{j : W_j \geq T\}| + q^{-1}}\right]$$

$$= \mathbb{E}\left[\frac{\left|\left\{j : W_j \geq T, \widehat{\text{sign}}_j \neq \text{sign}((\mathbf{X}_j - \widetilde{\mathbf{X}}_j)^\top \boldsymbol{\mu})\right\}\right|}{1 + |\{j : W_j \leq -T\}|} \cdot \frac{1 + |\{j : W_j \leq -T\}|}{|\{j : W_j \geq T\}| + q^{-1}}\right]$$

$$\leq q \cdot \mathbb{E}\left[\frac{\left|\left\{j : W_j \geq T, \widehat{\text{sign}}_j \neq \text{sign}((\mathbf{X}_j - \widetilde{\mathbf{X}}_j)^\top \boldsymbol{\mu})\right\}\right|}{1 + |\{j : W_j \leq -T\}|}\right]$$

since $|\{j : W_j \leq -T\}| \leq q \cdot |\{j : W_j \geq T\}|$ by definition of $T$. Now rewrite the set of sign errors as

$$\left\{j : W_j \geq T, \widehat{\text{sign}}_j \neq \text{sign}((\mathbf{X}_j - \widetilde{\mathbf{X}}_j)^\top \boldsymbol{\mu})\right\} = \left\{j : W_j \geq T, \text{sign}\left((\mathbf{X}_j - \widetilde{\mathbf{X}}_j)^\top \mathbf{y}\right) \neq \text{sign}((\mathbf{X}_j - \widetilde{\mathbf{X}}_j)^\top \boldsymbol{\mu})\right\}$$

$$= \left\{j : W_j \geq T, S_j \neq \text{sign}\left((\mathbf{X}_j - \widetilde{\mathbf{X}}_j)^\top \boldsymbol{\mu}\right)\right\},$$

where the first equality follows from the definition of $\widehat{\text{sign}}_j$ in (15), and the second from the definition $S_j = \text{sign}\left((\mathbf{X}_j - \widetilde{\mathbf{X}}_j)^\top \mathbf{y}\right) \cdot \text{sign}(W_j)$ as in Lemma 2 (since $\text{sign}(W_j) = +1$ for all $j$ in this set). Define $\widehat{\mathcal{H}}_0 = \{j : S_j \neq \text{sign}\left((\mathbf{X}_j - \widetilde{\mathbf{X}}_j)^\top \boldsymbol{\mu}\right)\}$ and let $\mathcal{V}$ be as in Lemma 2. By the tower law we obtain

$$\text{mFDR}_{\text{dir}} \leq q \cdot \mathbb{E}\left[\mathbb{E}\left[\frac{\left|\left\{j : W_j \geq T, S_j \neq \text{sign}\left((\mathbf{X}_j - \widetilde{\mathbf{X}}_j)^\top \boldsymbol{\mu}\right)\right\}\right|}{1 + |\{j : W_j \leq -T\}|} \;\middle|\; \mathcal{V}\right]\right]$$

$$= q \cdot \mathbb{E}\left[\mathbb{E}\left[\frac{\left|\left\{j \in \widehat{\mathcal{H}}_0 : W_j \geq T\right\}\right|}{1 + |\{j : W_j \leq -T\}|} \;\middle|\; \mathcal{V}\right]\right]$$

$$\leq q \cdot \mathbb{E}\left[\mathbb{E}\left[\frac{\left|\left\{j \in \widehat{\mathcal{H}}_0 : W_j \geq T\right\}\right|}{1 + \left|\left\{j \in \widehat{\mathcal{H}}_0 : W_j \leq -T\right\}\right|} \;\middle|\; \mathcal{V}\right]\right] .$$

Next, note that $\widehat{\mathcal{H}}_0 \in \mathcal{V}$ since $\mathbf{S} \in \mathcal{V}$ by Lemma 2, and so inside of the conditional expectation, we can treat the set $\widehat{\mathcal{H}}_0$ as fixed. Lemma 1 proves that conditional on $\mathcal{V}$, the signs of $\mathbf{W}$ are independent, with $\mathbb{P}\{\text{sign}(W_j) = -1 \mid \mathcal{V}\} \geq 1/2$ for all $j \in \widehat{\mathcal{H}}_0$. Therefore, applying Corollary 1 with $\rho = 1/2$ gives

$$\mathbb{E}\left[\frac{\left|\left\{j \in \widehat{\mathcal{H}}_0 : W_j \geq T\right\}\right|}{1 + \left|\left\{j \in \widehat{\mathcal{H}}_0 : W_j \leq -T\right\}\right|} \;\middle|\; \mathcal{V}\right] \leq 1,$$

which concludes the argument.

The proof for $\text{FDR}_{\text{dir}}$, when using knockoff+ instead of knockoff, follows similarly: we have

$$\text{FDR}_{\text{dir}} = \mathbb{E}\left[\frac{\left|\left\{j : W_j \geq T_+, \widehat{\text{sign}}_j \neq \text{sign}((\mathbf{X}_j - \widetilde{\mathbf{X}}_j)^\top \boldsymbol{\mu})\right\}\right|}{1 + |\{j : W_j \leq -T_+\}|} \cdot \frac{1 + |\{j : W_j \leq -T_+\}|}{1 \vee |\{j : W_j \geq T_+\}|}\right]$$

$$\leq q \cdot \mathbb{E}\left[\frac{\left|\left\{j : W_j \geq T_+, \widehat{\text{sign}}_j \neq \text{sign}((\mathbf{X}_j - \widetilde{\mathbf{X}}_j)^\top \boldsymbol{\mu})\right\}\right|}{1 + |\{j : W_j \leq -T_+\}|}\right] ,$$

where the last step uses the (slightly more conservative) definition of the threshold $T_+$ for the knockoff+ procedure. From then on, the argument is the same as before.

### B.3 Proof of Lemma 1

We first give a slight generalization of Lemma 4 in [1]:

**Lemma 3.** *Suppose that $B_1, \ldots, B_n \overset{\text{iid}}{\sim} \text{Bernoulli}(\rho)$. Let $J$ be a stopping time in reverse time with respect to the filtration $\{\mathcal{F}_j\}$, where $\mathcal{F}_j \ni B_1 + \cdots + B_j, B_{j+1}, \ldots, B_n$, and where the variables $B_1, \ldots, B_j$ are exchangeable with respect to $\mathcal{F}_j$. Then*

$$\mathbb{E}\left[\frac{1+J}{1 + B_1 + \cdots + B_J}\right] \leq \rho^{-1} .$$

*Proof of Lemma 3.* Define the sum

$$S_j = B_1 + \cdots + B_j \in \mathcal{F}_j ,$$

and define the process

$$M_j = \frac{1+j}{1 + B_1 + \cdots + B_j} = \frac{1+j}{1 + S_j} \in \mathcal{F}_j .$$

In [1] it is shown that $\mathbb{E}[M_n] \leq \rho^{-1}$, therefore, by the optional stopping time theorem it suffices to show that $\{M_j\}$ is a supermartingale with respect to $\{\mathcal{F}_j\}$. First, since $\{B_1, \ldots, B_{j+1}\}$ are exchangeable with respect to $\mathcal{F}_{j+1}$, we have

$$\mathbb{P}\{B_{j+1} = 1 \mid \mathcal{F}_{j+1}\} = \frac{S_{j+1}}{1+j} \,.$$

Therefore, if $S_{j+1} > 0$,

$$\mathbb{E}[M_j \mid \mathcal{F}_{j+1}] = \frac{1+j}{1+S_{j+1}} \cdot \mathbb{P}\{B_{j+1} = 0 \mid \mathcal{F}_{j+1}, S_{j+1}\} + \frac{1+j}{1+S_{j+1}-1} \cdot \mathbb{P}\{B_{j+1} = 1 \mid \mathcal{F}_{j+1}, S_{j+1}\}$$

$$= \frac{1+j}{1+S_{j+1}} \cdot \frac{1+j-S_{j+1}}{1+j} + \frac{1+j}{1+S_{j+1}-1} \cdot \frac{S_{j+1}}{1+j}$$

$$= \frac{1+j-S_{j+1}}{1+S_{j+1}} + 1 = \frac{1+(j+1)}{1+S_{j+1}} = M_{j+1} \,.$$

If instead $S_{j+1} = 0$, then trivially $S_j = 0$ also, and so $M_j = 1 + j < 2 + j = M_{j+1}$. This proves that $\{M_j\}$ is a supermartingale with respect to $\{\mathcal{F}_j\}$, as desired. $\square$

**Corollary 2.** *Suppose that $\mathcal{A} \subseteq [n]$ is fixed, while $B_1, \ldots, B_n \overset{\text{iid}}{\sim}$ Bernoulli$(\rho)$. Let $J$ be a stopping time in reverse time with respect to the filtration $\{\mathcal{F}_j\}$, where $\mathcal{F}_j \ni \sum_{i \leq j, i \in \mathcal{A}} B_i$, and the variables $\{B_i : i \leq j, i \in \mathcal{A}\}$ are exchangeable with respect to $\mathcal{F}_j$. Then*

$$\mathbb{E}\left[\frac{1 + |\{i \leq J : i \in \mathcal{A}\}|}{1 + \sum_{i \leq J, i \in \mathcal{A}} B_i}\right] \leq \rho^{-1} \,.$$

*Proof of Corollary 2.* Let $\mathcal{A} = \{i_1, \ldots, i_m\}$ where $1 \leq i_1 < \cdots < i_m \leq n$. Then by considering the i.i.d. sequence

$$B_{i_1}, \ldots, B_{i_m}$$

in place of $B_1, \ldots, B_n$, we see that this result is equivalent to Lemma 3. $\square$

*Proof of Lemma 1.* We may assume $\rho < 1$ to avoid the trivial case. We now create a different construction for the $B_i$'s. First, generate a random set $\mathcal{A} \subseteq [n]$ where for each $i$, independently,

$$\mathbb{P}\{i \in \mathcal{A}\} = \frac{1 - \rho_i}{1 - \rho} \,.$$

Next, define variables $Q_1, \ldots, Q_n \overset{\text{iid}}{\sim}$ Bernoulli$(\rho)$, which are generated independently of the random set $\mathcal{A}$. Finally, define

$$B_i = Q_i \cdot \mathbb{1}_{i \in \mathcal{A}} + \mathbb{1}_{i \notin \mathcal{A}} \,. \tag{31}$$

Then, clearly, the $B_i$'s are mutually independent with $\mathbb{P}\{B_i = 1\} = \rho_i$, as required by the lemma. Next, since $B_i = Q_i \cdot \mathbb{1}_{i \in \mathcal{A}} + \mathbb{1}_{i \notin \mathcal{A}}$ for all $i$, we have

$$\frac{1+J}{1 + B_1 + \cdots + B_J} = \frac{1 + |\{i \leq J : i \in \mathcal{A}\}| + |\{i \leq J : i \notin \mathcal{A}\}|}{1 + \sum_{i \leq J, i \in \mathcal{A}} Q_i + |\{i \leq J : i \notin \mathcal{A}\}|} \leq \frac{1 + |\{i \leq J : i \in \mathcal{A}\}|}{1 + \sum_{i \leq J, i \in \mathcal{A}} Q_i} \,,$$

where the last step uses the identity $\frac{a+c}{b+c} \leq \frac{a}{b}$ whenever $0 < b \leq a$ and $c \geq 0$. Therefore, it will be sufficient to bound the right-hand side. We will use Corollary 2 to prove that

$$\mathbb{E}\left[\frac{1 + |\{i \leq J : i \in \mathcal{A}\}|}{1 + \sum_{i \leq J, i \in \mathcal{A}} Q_i} \,\middle|\, \mathcal{A}\right] \leq \rho^{-1} \,, \tag{32}$$

which will be sufficient to prove the lemma via the tower law of expectations.

To prove (32), first let $\widetilde{Q}_i = Q_i \cdot \mathbb{1}_{i \in \mathcal{A}}$, and define a filtration $\{\mathcal{F}'_j\}$ where $\mathcal{F}'_j$ is the $\sigma$-algebra generated as

$$\mathcal{F}'_j = \sigma\left(\left\{\widetilde{Q}_1 + \cdots + \widetilde{Q}_j, \widetilde{Q}_{j+1}, \ldots, \widetilde{Q}_n, \mathcal{A}\right\}\right).$$

Next, for any $j$, by (31) we see that

$$B_1 + \cdots + B_j, B_{j+1}, \ldots, B_n \in \mathcal{F}'_j \quad \Rightarrow \quad \mathcal{F}'_j \supseteq \mathcal{F}_j,$$

and so $T$ is a stopping time (in reverse time) with respect to $\{\mathcal{F}'_j\}$ also. Next, the variables $\{Q_i : i \leq j, i \in \mathcal{A}\}$ are trivially exchangeable with respect to $\mathcal{F}'_j$ (since the $Q_i$'s are i.i.d., and independent from $\mathcal{A}$). Finally, since the $Q_i$'s are independent from $\mathcal{A}$, the desired bound (32) follows directly from Corollary 2 after conditioning on $\mathcal{A}$. $\square$

## C  Detailed results for GWAS experiment

In this section we present the full results obtained in the GWAS experiment (see Section 6.2 in the main paper for details). Table 2 gives the results for the LDL phenotype, and Table 3 gives the results for the HDL phenotype.

31 02/2016; Revised 09/2017 and 05/2018| Selection frequency (by region) | Selection frequency (by SNP) | SNP name | Chromosome | Base pair position | Nearby SNPs found in Willer et al. [36] | Nearby SNPs found in Sabatti et al. [26] |
|---|---|---|---|---|---|---|
| 9 | 7 | rs693 | 2 | 21232195 | rs1367117 | rs693 |
|   | 4 | rs3923037 |   | 21148274 |   |   |
|   | 4 | rs754524 |   | 21311541 |   |   |
|   | 2 | rs6754295 |   | 21206183 |   |   |
|   | 1 | rs1429974 |   | 21300770 |   |   |
| 7 | 7 | rs646776 | 1 | 109818530 | rs629301 | rs646776 |
| 7 | 7 | rs4844614 | 1 | 207875175 |   | rs4844614 |
| 7 | 7 | rs157580 | 19 | 45395266 | rs4420638 | rs157580 |
| 7 | 5 | rs11668477 | 19 | 11195030 | rs6511720 | rs11668477 |
|   | 2 | rs11878377 |   | 10965270 |   |   |
|   | 2 | rs1541596 |   | 10987013 |   |   |
|   | 1 | rs10409243 |   | 10332988 |   |   |
|   | 1 | rs688 |   | 11227602 |   |   |
| 6 | 2 | rs174556 | 11 | 61580635 | rs174546 | rs174537 |
|   | 2 | rs174450 |   | 61641542 |   | rs102275 |
|   | 1 | rs579383 |   | 61536583 |   | rs174546 |
|   | 1 | rs1535 |   | 61597972 |   | rs174556 |
|   |   |   |   |   |   | rs1535 |
| 4 | 4 | rs2728487 | 7 | 47209007 |   |   |
| 4 | 4 | rs9696070 | 9 | 89230779 |   |   |
| 3 | 3 | rs1342165 | 1 | 34800325 |   |   |
| 3 | 3 | rs557435 | 1 | 55520864 | rs2479409 |   |
| 3 | 3 | rs4906908 | 15 | 27040082 |   |   |
| 3 | 2 | rs579163 | 11 | 116713630 | rs964184 |   |
|   | 1 | rs518181 |   | 116772787 |   |   |
|   | 1 | rs11216267 |   | 116952392 |   |   |
| 2 | 2 | rs2802955 | 1 | 235015199 | rs514230 |   |
| 2 | 2 | rs10953541 | 7 | 107244545 |   |   |
| 2 | 2 | rs905502 | 8 | 3134810 |   |   |
| 2 | 2 | rs1897318 | 8 | 124864404 |   |   |
| 2 | 1 | rs10062361 | 5 | 74565153 | rs12916 |   |
|   | 1 | rs6896136 |   | 74787310 |   |   |
| 2 | 1 | rs12427378 | 12 | 51074199 |   |   |
|   | 1 | rs2139930 |   | 51089287 |   |   |
| 1 | 1 | rs7574918 | 2 | 165939179 |   |   |
| 1 | 1 | rs6749903 | 2 | 223653986 |   |   |
| 1 | 1 | rs3733262 | 4 | 129439663 |   |   |
| 1 | 1 | rs981862 | 5 | 93615466 |   |   |
| 1 | 1 | rs995124 | 7 | 41764992 |   |   |
| 1 | 1 | rs2351643 | 8 | 63076996 |   |   |
| 1 | 1 | rs945559 | 10 | 89823147 |   |   |
| 1 | 1 | rs1955105 | 12 | 115344782 |   |   |
| 1 | 1 | rs2456930 | 15 | 62687339 |   |   |
| 1 | 1 | rs4433842 | 17 | 64322683 | rs1801689 |   |
| 1 | 1 | rs4799847 | 18 | 33686203 |   |   |

**Table 2:** Results of GWAS experiment for the LDL phenotype (see Section 6.2 in the main paper for details and interpretation of the results). Physical reference positions for SNPs were drawn from Human Genome Build 37/HG19.

| Selection frequency (by region) | Selection frequency (by SNP) | SNP name | Chromosome | Base pair position | Nearby SNPs found in Willer et al. [36] | Nearby SNPs found in Sabatti et al. [26] |
|---|---|---|---|---|---|---|
| 9 | 9 | rs3764261 | 16 | 56993324 | rs3764261 | rs3764261 |
|   | 9 | rs7499892 |   | 57006590 |   |   |
| 8 | 8 | rs1532085 | 15 | 58683366 | rs1532085 | rs1532085 |
| 4 | 2 | rs2575875 | 9 | 107662494 | rs1883025 |   |
|   | 2 | rs2740486 |   | 107666513 |   |   |
| 4 | 2 | rs255049 | 16 | 68013471 | rs16942887 | rs255049 |
|   | 2 | rs255052 |   | 68024995 |   |   |
| 3 | 3 | rs7120118 | 11 | 47286290 | rs3136441 | rs2167079 |
|   |   |   |   |   |   | rs7120118 |
| 1 | 1 | rs12139970 | 1 | 230406460 | rs4846914 |   |
| 1 | 1 | rs6728178 | 2 | 21193946 |   |   |
| 1 | 1 | rs173738 | 5 | 16725880 |   |   |
| 1 | 1 | rs2817056 | 6 | 35734051 |   |   |
| 1 | 1 | rs2375016 | 12 | 109227914 | rs7134594 |   |

**Table 3:** Results of GWAS experiment for the HDL phenotype (see Section 6.2 in the main paper for details and interpretation of the results). Physical reference positions for SNPs were drawn from Human Genome Build 37/HG19.